\documentclass{aa}  

\usepackage{graphicx}
\usepackage{txfonts}
\usepackage{lipsum}
\usepackage{subcaption}         % necessary for continued figures, example in section 3
\usepackage{lscape}             % to rotate a single page table, example in appendix.
\usepackage{placeins}           % useful with \FloatBarrier, to keep 
\usepackage{footmisc}
\usepackage{float}

\usepackage{hyperref}
\begin{document}

   \title{A multi-band radio flux density catalog of ICRF3 sources using the Onsala Twin Telescopes}

   \author{Alva  Kinman\inst{1}\fnmsep\thanks{Corresponding author: kinman@chalmers.se}
        \and Rüdiger Haas\inst{1}
        \and Karine Le Bail\inst{1}
        \and Jun Yang\inst{1} 
        \and Eskil Varenius\inst{2}
        }

   \institute{Department of Physics and Astronomy, Chalmers University of Technology, Onsala Space Observatory SE-43992 Onsala, Sweden
   \and on the leave from Chalmers University of Technology, Onsala Space Observatory SE-43992 Onsala, Sweden}

   \date{}

  \abstract
   {The VLBI Global Observing System (VGOS) is the next generation system for geodetic and astrometric very long baseline interferometry (VLBI). 
   VGOS is broadband and uses fast-moving antennas that are capable of observing more scans per minute than its predecessors. 
   To optimize the observing time for each source in geodetic schedules, a flux density catalog is needed for the sources that are observed at the VGOS frequencies.}
   {The aim of this work is to monitor the flux densities of geodetic sources in the VGOS bands. 
The obtained flux density time series can be used for more effective scheduling of geodetic and astrometric VLBI experiments, as well as probing active galactic nuclei (AGN) physics.} 
   {The Onsala Twin Telescopes have been used as a single baseline interferometer to measure flux densities of AGN that are part of the International Celestial Reference System (ICRF3). 
   The telescopes observed four frequency bands simultaneously, centered at 3.2, 5.5, 6.6 and 10.4~GHz.
   Both locally planned flux monitoring sessions and international geodetic experiments have been analyzed. 
   The data have been correlated using the Distributed FX (DiFX) software correlator and calibrated using the Common Astronomy Software Applications (CASA). 
   The possibility of predicting geodetic signal-to-noise ratios (S/N) using the measured flux densities was also tested.}
   {Simultaneous light curves in up to four frequencies have been obtained for 361 sources, of which 270 sources have at least five observation epochs. 
   The majority of the sources vary significantly in flux density during the measurement period. Most sources have a flat or inverted spectrum, with only 6 \% having a steep spectrum. There is also evidence of spectral index variation over time for a number of sources.
   Furthermore, the flux densities from this work were shown to more precisely predict geodetic signal-to-noise ratios compared to the standard VGOS flux density catalog, in particular when considering the most variable sources.}
   {The majority of the radio sources observed by VGOS show significant variation in flux density.
   Such variation needs to be taken into account to obtain the most optimal VGOS schedules.
   The flux density catalog presented here is expected to be of use for both astronomy and geodesy. 
   We plan to continue the monitoring program, potentially adding more sources and more participating stations.}

   \keywords{Catalogs -- Galaxies: active --                Radio continuum: galaxies -- 
                Instrumentation: interferometers -- Reference systems
               }

   \maketitle

    \nolinenumbers

\section{Introduction}

Active galactic nuclei (AGN) are often variable at radio frequencies. 
The emission from these objects is powered by a supermassive black hole located in the center of a galaxy. 
Gas falling in towards the black hole is heated and ionized, forming an accretion disk. 
Magnetic fields in the disk can cause plasma jets to be launched outwards at relativistic speeds \citep[see review by][]{blandford2019}. 
The emission observed in the radio regime mainly consists of synchrotron radiation emitted by the high energy electrons in the plasma \citep{padovani2017}. 
Active galactic nuclei are dynamic objects that may vary in both brightness and morphology, commonly due to shocks forming in the jet (\citealp[e.g.][]{savolainen2002,hovatta2008}). 
In the case of quasars and blazars, the observed flux density is additionally boosted by relativistic beaming \citep{urry1995}, which depends on the angle between the jet and the line of sight. 
Observed flux density variations can thus both be caused by intrinsic luminosity changes and by changes in the viewing angle, which may arise from jet precession \citep[][and references therein]{britzen2023}.

The variability of AGN is not only of interest for astronomers, but also has consequences for the field of geodetic very long baseline interferometry (VLBI). 
VLBI is a technique in which multiple radio telescopes around the world observe the same radio source simultaneously, measuring the difference in arrival time of the wavefront to each telescope  \citep[see reviews by][]{sovers1998,schuh2012}. 
Geodetic VLBI is used to precisely measure telescope positions, contributing to the International Terrestrial Reference Frame \cite[ITRF,][]{altamimi2023}. 
The technique also provides Earth orientation parameters (EOP), which describe the orientation of the Earth in space. 
The EOP provide the connection between the ITRF and the International Celestial Reference Frame \citep[ICRF3,][]{charlot2020}. 
The targets of geodetic VLBI observations are bright AGN, typically quasars and blazars.

Since 1979, geodetic VLBI observations have been carried out in the S and X band (2.2–2.4 GHz and 8.2–8.9 GHz, respectively) with the so-called legacy S/X system. 
To meet increased demands for accuracy put forth by the Global Geodetic Observing System (GGOS), the International VLBI Service for Geodesy and Astrometry \citep[IVS,][]{nothnagel2017} started the design of a next generation system for geodetic VLBI \citep{behrend2009}.

This new system, known today as the VLBI Global Observing System (VGOS), has been in operation since 2020 \citep{behrend2023}. 
The aim of VGOS is to improve
 the accuracy of geodetic parameters, with the goal of reaching 1~mm for station positions and 0.1~mm/yr for station velocities \citep{petrachenko2009}. 
VGOS was designed to observe in the frequency range of 2--14~GHz. Currently, only the 3--11~GHz range is used operationally since some stations are not yet capable of observing the full range.
Within this range, four frequency bands are placed, each consisting of eight 32~MHz wide sub-bands. 
The large spanned bandwidth allows for precise estimation of the group delay.

VGOS antennas were built to be small and fast-moving in order to address one of the most important error sources in geodetic VLBI: the troposphere. 
The troposphere introduces time delays corresponding to several meters, which must be estimated and removed to achieve precise positions \citep{sovers1998}. 
To accurately track the tropospheric delays, these estimates need to be done at sufficiently short time intervals.
This requires a large number of observations in each such time interval, so that the sky above each antenna is well-sampled. 

The number of observations in a fixed time interval is limited by the time that the telescopes spend observing each source. 
The on-source time $t$ required for a successful observation is in turn limited by the sensitivity of the telescopes, the desired signal-to-noise ratio (S/N) and the source flux density \citep[][Eq. (1.7) and (6.48)]{thompson2017}
\begin{equation}
\label{eq:obstime}
    t = \left( \frac{S/N}{\eta F_\nu}\right)^2 \frac{\rm SEFD_1\cdot SEFD_2}{2\Delta\nu},
\end{equation}
where $F_\nu$ is the source flux density, $\rm SEFD_{1,2}$ are the system equivalent flux densities of the telescopes, $\Delta \nu$ is the recorded bandwidth and $\eta$ represents losses from digitization and correlation. 
Consequently, a short observation time is enough for a bright source, while a fainter source requires a longer observation duration to get a detection.

Flux densities of geodetic sources in the S and X band, used for legacy S/X observations, are available and updated monthly\footnote{\url{https://github.com/nvi-inc/sked_catalogs}} \citep{gipson2016}. They are estimated by solving for the flux density in Eq. \eqref{eq:obstime}, inserting observed S/N values and fixed SEFD values for each station. 
When scheduling new experiments, the observing time is adapted for each source to reach the target S/N.
However, there is so far no equivalent flux density catalog available for VGOS. 

\cite{schartner2025} investigated VGOS scheduling based on flux density in the VGOS bands.
Flux densities were derived from observed S/N and station SEFDs in previous VGOS-Operations (VO) sessions. 
Their scheduling method was able to increase the number of scans compared to the previous method with fixed 30-second scan durations, but there were some discrepancies between predicted and measured S/Ns. 
Furthermore, this approach accumulated data from several months to estimate a single flux density value per source and baseline. 
The catalog is based on observations from 2022 and not regularly updated. 
The current flux density of the source can therefore differ from the value in the catalog, as compact AGN are known to be variable on timescales of hours to years \citep[e.g.][]{wagner1995,barvainis2005,hovatta2008,richards2011,raiteri2025}.

With the aim of tracking variations in AGN flux densities for use in geodetic scheduling, we have devised a flux monitoring program at Onsala Space Observatory (OSO). 
The Onsala Twin telescopes (OTT), which are part of VGOS, are used as a single baseline interferometer. 
Flux densities are measured in the four operationally used VGOS bands, 
known as VGOS band A, B , C and D. 
Flux monitoring sessions are scheduled approximately once per month. 
The work builds on the pilot study by \cite{varenius2022}, which demonstrated that the OTT could be successfully used for flux monitoring. 
Though geodetic VLBI is the primary motivator for this work, we also envision a dual use of this dataset, as multi-band light curves are a valuable probe of the physical processes inside AGN \citep[e.g.][]{kudryavtseva2011,hovatta2019,hsu2023}. 

Section \ref{sec:method} presents the observations and data analysis. Section \ref{sec:results} introduces the flux density catalog and explores some of its properties. 
Section \ref{sec:snr} presents a comparison between theoretical and observed S/N in geodetic experiments based on the flux densities from the catalog.  Section \ref{sec:discussion} contains a discussion of the results as well as future outlook. Finally, the conclusions are presented in Section \ref{sec:conclusions}.

\section{Observations and analysis}
\label{sec:method}

The experimental setup as well as the complete analysis procedure is described below. 
Both observations and analysis methods bear similarities to \cite{kinman2025}, but there are significant adjustments in the analysis.

\subsection{Single-baseline flux density observations}

The experiments analyzed in this study were carried out with the OTT, consisting of ONSA13NE (Oe) and ONSA13SW (Ow).
The telescopes have a diameter of 13.2~m each and are separated by 75~m.
The telescope feeds receive two orthogonal linear polarizations, here denoted X and Y.

Two types of experiments are analyzed in this work. 
The first type is dedicated flux monitoring (FM) sessions, planned and performed locally at OSO. 
The second type is VGOS-Operational (VO) sessions coordinated by the IVS where a global network of stations participated. 
In this work however, only data from the OTT baseline were used.

\begin{figure*}[t!]
    \centering
    \includegraphics[width=0.9\linewidth]{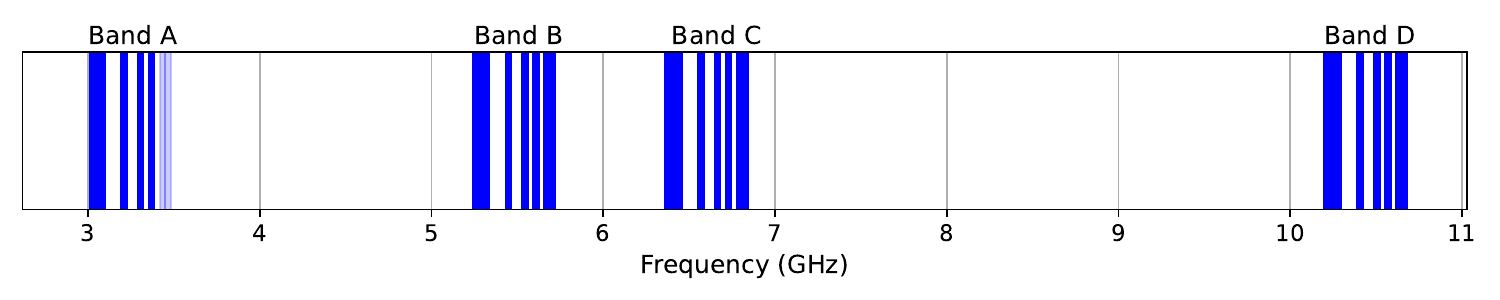}
    \caption{Operational VGOS frequency setup. Each band is 480~MHz wide in total and contains eight sub-bands with a bandwidth of 32~MHz. The two highest sub-bands in band~A (shown in a lighter shade) were excluded from the analysis due to strong radio-frequency interference.}
    \label{fig:vgos-setup}
\end{figure*}

For both types of experiments, the standard VGOS frequency setup was used. 
It currently consists of four bands centered at 3.2, 5.5, 6.5 and 10.4~GHz, each containing eight 32~MHz wide sub-bands. 
The complete setup is shown in Fig.~\ref{fig:vgos-setup} and numerically presented in e.g. \cite{varenius2022}. 
The two highest sub-bands in band~A were excluded due to persistent disturbing electromagnetic radiation, often referred to as radio-frequency interference (RFI).

The VLBI Field System \citep[FS,][]{himwich2000} was used for controlling the antennas and recording system temperatures during the observations.
The FS version was 10.1 until May 2024 and 10.2 onward.

\subsubsection{Local flux monitoring experiments}

The first FM experiments were performed in 2021 and are described in \cite{varenius2022}. 
The current flux monitoring program started in January 2023 and is still ongoing. In this paper we report results from January 2023 until May 2026.

The FM experiments target all the defining sources of the ICRF3 that are visible from OSO at above ten degrees elevation, resulting in a catalog of 209 sources. 
Six additional AGN, also part of the ICRF3 catalog but not defining, have been added to the catalog since they may be of astronomical interest: 
the supermassive binary black hole candidates OJ287 \citep{sillanpaa1988}, 1044+719 \citep{kun2022}, and 1928+738 \citep{roland2015}, as well as the candidate neutrino sources 1424+240, 1542+616, and 0609+413 \citep{aartsen2020,abbasi2026}. 
In addition, three flux density calibrator sources are observed in each experiment: 3C147, 3C286, and 3C295.

The FM experiments are scheduled using the software VieSched++ \citep{schartner2019}. The experiment duration is set to 26 hours, and during this time as many sources as possible are observed. Each source is observed in three fixed-duration scans. For sources brighter than 0.6~Jy in the previous experiment, the duration is 60~seconds, while flux density calibrators and fainter sources are observed for 120~seconds.

The experiments are carried out approximately once per month. 
Longer gaps in the time series are caused by technical issues with the telescopes and lack of available data storage space.

\subsubsection{IVS VGOS Operational sessions}

Data from the IVS VO sessions were downloaded from the Crustal Dynamics Data Information Systems (CDDIS)\footnote{\url{https://cddis.nasa.gov/archive/vlbi/ivsdata/swin/}} database. 
Since these experiments lack observations of flux density calibrators, flux calibration scans have been added locally at OSO after each VO session. 
These consist of one 120 second long scan each towards one or more of the sources 3C147, 3C286, and 3C295. 
The VO experiments available to our analysis are limited by a few factors. Firstly, both Oe and Ow need to have participated in the session and be free from significant technical issues. 
Secondly, the correlated data need to be available on CDDIS, which is not the case for all sessions. 
In total, we present results from 26 VO sessions in this work.

The VO sessions have a duration of 24 hours and the number of scans is different for each source, but typically larger than in the flux monitoring experiments. 
Until mid-2024, all scans had a duration of 30 seconds. 
In the more recent experiments, the scan duration is determined by the approximated flux density from S/X catalogs and/or \cite{schartner2025} and varies between 10 and 30 seconds. 
The source catalogs used for these experiments contain some ICRF3 defining sources, but also many non-defining sources.
Our list of analyzed sources is thus extended by the addition of the VO sessions.
Note that the list of observed sources differs significantly between experiments. 
A source is observed, on average, in 8 out of 26 analyzed VO experiments. 
Many new sources were added to the VO experiments in 2026, resulting in a number of sources that only have one or two epochs in the current version of the flux density catalog. 
The catalog will be continuously updated as new observations become available.

\subsection{Correlation}

The FM sessions were correlated at OSO using DiFX \citep{deller2007} and all four correlation products XX, XY, YX, and YY were produced. During the three years we used DiFX versions 2.5.5, 2.6.3, and 2.8.1. We have verified that the visibilities produced by the three versions agree within 0.1~\%.
The flux calibration scans observed in connection to the VO sessions were correlated in the same way as the FM sessions. 
The spectral resolution was set to 100~kHz.
A high spectral resolution is advantageous since there is a need to filter out the phase calibration peaks from the cross-power spectrum. 
The phase calibration system inserts a tone into the signal path that is later used to calibrate instrumental delays in geodetic analysis. 
Since Oe and Ow get their time and frequency reference from the same hydrogen maser, their phase calibration signals are highly correlated. 
The phase calibration signals appear as tall, narrow peaks in the spectra. 
A high spectral resolution allows these peaks to be removed while keeping as much of the bandwidth as possible. Note that we for that reason used a spectral resolution of 100~kHz for the VO calibration scans as well. The VO sessions are typically correlated at a coarser spectral resolution.

\subsection{Calibration in CASA}

After correlation, the data were converted to FITS-IDI format \citep{greisen2011}. 
An antab file was also generated, containing the gain curve and system temperature data from the experiment. 
A~flat gain curve was assumed for the OTT, since the small telescopes are not expected to be subject to large gravitational deformation. However, this assumption could be refined in the near future (see Section \ref{dis:errors}). 
The data were then processed using the Common Astronomy Software Applications \citep[CASA,][]{casateam2022} version 6.7, together with VLBI tools developed by JIVE\footnote{\url{https://github.com/jive-vlbi/casa-vlbi}}.

\subsubsection{Main pipeline}
A CASA pipeline was created to perform calibration of the visibilities. 
The pipeline is a further development of the pipeline used in \cite{varenius2022}. 
The code is available on GitHub\footnote{\url{https://github.com/alvakinman/OSO_flux_monitoring}}.

Firstly, the visibilities and system temperature values were imported into a CASA measurement set. 
The script \texttt{gc.py} from JIVE was used to import the gain information. 
Then, the frequency bins containing peaks from the phase calibration signal were flagged. 
The lower 10~\% and upper 5~\% of the  frequency bins in each sub-band were also flagged. 
These bins have a low signal-to-noise ratio and may negatively impact the bandpass normalization if they are kept. 
After that, outliers were flagged using the algorithm \textit{rflag}.

After initial flagging, amplitude calibration tables were created. 
They include a gain table, a system temperature table, and an autocorrelation correction table. 
The autocorrelation correction compensates for sub-optimally set sampler levels in the data recording. 
While some correlators already apply this correction, DiFX does not \citep{vanbemmel2022}.

Next, the data were fringe-fitted using the CASA task \textit{fringefit}. 
First, instrumental fringe-fitting was performed using one scan of a bright source. 
Then a multi-band fringe-fitting, with all bands combined, was performed on a few bright scans. 
In a typical VLBI experiment, the change in delay and visibility phase over time is dominated by atmospheric effects. 
Fringe-fitting is thus needed for each scan, since the difference in atmospheric delay is both time- and direction dependent. 
However, the OTT are located only 75 meters apart, which makes the difference in atmospheric delay negligible. 
Instead, the change in phase and delay throughout the experiment is dominated by clock rate errors in the correlation. 
This results in a smooth, approximately linear phase and delay evolution, which means that it is possible to run \textit{fringefit} on a small number of scans and interpolate between them. 

For each experiment, a list of sources were picked to be used in multi-band fringe fitting.
For the FM experiments, the scans of the flux calibrators, OJ287, 0133+476, 0552+398, and 0059+581 were used by default. 
These are known to be bright (several Jy). 
For the VO experiments, there are hourly 120-second scans of bright sources that can be used for this purpose. 
In addition, the first and last scan of the experiment was included, to allow phase and delay rates to be interpolated also in the beginning and end of the experiment. 
Additional sources were added to the fringe-fitting list for some experiments to make sure the phase drift between adjacent \textit{fringefit} solutions did not exceed 180 degrees. 
Such a large drift would cause the interpolation to fail when the solution is applied.

After fringe fitting, a bandpass calibration was done on a single scan of a flux calibrator source (including both amplitude and phase corrections), after which all calibration tables were applied to the dataset. Lastly, a second round of the \textit{rflag} algorithm was run.

The statistical weights of the visibilities were updated by the CASA task \textit{statwt}, based on the standard deviation of the visibilities in each sub-band and 10-second time window. Then, a weighted average was obtained over time and frequency within each sub-band. 
The amplitudes of the XX, YY, XY, and YX correlation products were then averaged over all scans of the source. 
For each source, we obtain data from 32 sub-bands: in each sub-band we obtain an XX, XY, YX, and YY amplitude. 
We also save the uncertainties of these values, based on the aforementioned statistical weights.

Given these correlation amplitudes, it is possible to calculate a Stokes I value:
\begin{equation}
    I = \frac{XX+YY}{2}.
\end{equation}
No parallactic angle correction was necessary due to the short baseline between Oe and Ow.

\subsubsection{Special considerations for VO}

The above described calibration process was used for the dedicated FM experiments. 
The calibration process for the VO experiments had a few key differences.

Firstly, correlated data from the VO sessions were downloaded from CDDIS. 
However, flux calibration scans were not present in that dataset, since they were not part of the IVS schedule. 
Instead, flux calibration scans were observed with the OTT directly after the VO sessions, and then correlated locally at OSO. 
As a consequence, delays and delay rates were often different between the VO dataset and the calibrator scans. Therefore, the flux calibration scans were treated as a separate dataset during the calibration. 
However, it would be inaccurate to use different bandpass solutions for the calibrators and the main dataset. 
The bandpass solution may affect the scaling factors obtained from the calibration (see Section \ref{met:fluxcal}), and thus it is most accurate to use the same bandpass solution for every scan. 
Therefore, the bandpass solution obtained from the calibrator dataset was applied to the VO dataset as well.

Secondly, the VO data contain many short scans of the same source. 
The large number of scans provides an opportunity for outlier filtering. 
After calibration, the median and median absolute deviation (MAD) of the amplitude of all scans to each source was computed. 
Any scan that deviated more than 5~MAD from the median was removed. 
This approach would be less useful for the FM experiments, given that each source is only observed three times.

\subsection{Absolute flux density calibration}

\label{met:fluxcal}
As a final step in the calibration for both FM and VO experiments, the three flux calibrator sources 3C147, 3C286, and 3C295 were used for absolute flux density calibration. 
The expected flux density of each calibration source at each sub-band was obtained from the flux models presented in \citet{perley2017}. 
To calibrate source flux densities, the expected flux density was divided by the measured Stokes I value of the calibrators in each sub-band. 
The obtained scale factors were averaged over the three calibrator sources. 
The scale factor in each sub-band was then used to correct the amplitudes of all sources.
The scale factors were calculated separately for each experiment, to account for instrumental variation over time.
The scale factor also corrects systematic gain differences between sub-bands.

If the flux calibrator scans are affected by disturbing signals, it could change the scale factor. To avoid using incorrect scale factors, we flagged any sub-band where the scale factor was larger than 2, smaller than 0.5 or where any single flux calibrator had a scale factor that differed from the mean by more than 20~\%.

\subsection{Removal of outliers}

\label{met:outliers}
The OTT baseline is heavily affected by RFI in some sub-bands.
It appears mainly in VGOS band A and the lower part of band B. 
This is to some degree mitigated by the automatic flagging algorithm in the CASA pipeline. 
However, the automatic flagging fails to catch broad-band RFI. 
Therefore, the RFI-polluted flux density values need to be identified and filtered out after the CASA processing.

In \cite{varenius2022} and \cite{kinman2025}, filtering was done based on the cross-polarization amplitudes. 
Under the assumption that the radio source is unpolarized, 
the cross-polarization amplitudes are close to zero. If a significant cross-polarization amplitude is detected, it could indicate the presence of polarized RFI signals. For example, signals used for radio communication are often polarized. 

Although the cross-polarization filtering proved effective in the previous works for removing bad data in band A, it comes with a few disadvantages. 
Firstly, the cross-polarization filtering tends to excessively flag faint sources, also in bands that are not significantly affected by RFI. 
Additionally, it flags VO data more severely than FM data. 
A discussion of the reason for this is presented in Section~\ref{dis:averaging}. 
Secondly, this approach hinders us from observing radio sources that are polarized. Several sources in our catalog are in fact polarized at radio frequencies, for example 0954+658 and 1717+178 \citep{pushkarev2023}. The previous method risks removing valid data for such sources.

With these disadvantages in mind, we here choose a different approach. 
The radio spectra of the sources are expected to be smooth, so that the flux density does not vary dramatically over one 480~MHz wide band. The flux density can be described by a power-law of the form $F_\nu \propto \nu^{-\alpha}$, where $\alpha$ is called the spectral index.
Considering that the spectral index of these sources is typically in the range of $-1$ to 1 (see Section \ref{res:properties}), the maximum expected flux variation over a band can be estimated by
\begin{equation}
    \frac{F_{\nu, max}}{F_{\nu,min}} = \left ( \frac{\nu_{high}}{\nu_{low}} \right )^{\pm 1},
\end{equation}
where $F_{\nu}$ is the flux density and $\nu_{high}, \nu_{low}$ are the frequencies of the highest and lowest sub-band respectively. Band A has the largest ratio between $\nu_{high}$ and $\nu_{low}$: 1.16. This means that the flux density difference between the highest and lowest sub-band is expected to be at most 16 \%, 
plus a measurement error on the order of a few percent. The flux density difference between the center of the band and the lowest/highest sub-band is approximately half of that. 
If some of the sub-bands are affected by RFI however, a larger variation typically arises. 
To get a robust catalog, we used a filtering procedure as described below.

First, we applied a 3$\sigma$ criterion: if the XX or YY amplitude was smaller than three times the corresponding uncertainty, the sub-band was removed to avoid false detections due to noise.
Next, we computed the median XX and YY amplitude in each band. If any sub-band deviated more than 15~\% from the median in either XX or YY, it was removed. 
If there were fewer than four sub-bands left after these filtering steps, we removed them as well. 
If the majority of the sub-bands are far from the median, the median itself is not reliable. 
We then computed the weighted average of the remaining sub-bands within each band. 

Just like the previously used cross-polarization criterion, this outlier removal process flags most of the data points in band~A. 
However, it keeps more data in the other bands, for fainter sources in particular. 
This is illustrated in Appendix~\ref{app:outlier}, Fig.~\ref{fig:failrate-fm}-\ref{fig:failrate-vo}.

\subsection{Error estimation}
\label{met:errors}
There are various errors to consider for these flux density measurements. 
Firstly, there is an individual uncertainty of each sub-band. 
This is comprised of mainly thermal noise and uncertainty in the scaling factor.
The theoretical thermal noise is given by
\begin{equation}
\label{eq:noise}
    \sigma = \frac{1}{\eta}\frac{\sqrt{\rm SEFD_1 \cdot SEFD_2}}{\sqrt{2\Delta\nu t}}.
\end{equation}
\citep[][Eq. (1.7) and (6.50)]{thompson2017}
The actual standard deviation of visibilities within a sub-band is usually close to what is predicted by this relation. 

Secondly, there is scatter among the sub-band flux densities within a band. 
This scatter is usually of the same order as what is expected from propagating the individual sub-band uncertainties, but can be larger or smaller in individual observations. 
The best agreement is found in band C and D, where the propagated uncertainties and the scatter-based uncertainties both have a median of 1~\% of the flux density. 
In band A, the scatter uncertainties are approximately 3~\% while the propagated uncertainties are 2~\%.

Thirdly, there are systematic errors caused by e.g. atmospheric opacity, the gain curve, and uncertainties in the reference flux densities of the calibrators.
These errors will typically be proportional to the flux density of the source, with a possible frequency dependence.
Additionally, RFI that is not caught by the outlier filter may introduce errors. The systematic errors are discussed further in Section~\ref{dis:errors}.

To estimate the uncertainties of our flux densities, we did the following.
First, we calculated 1) the propagated statistical error and 2) the standard deviation of flux densities in the band divided by the square root of the number of sub-bands. 
The largest of these errors was used and this value will henceforth be called the random error. 
Then we added in quadrature a three-percent uncertainty, to account for the systematic errors. 
To verify that this error size is realistic, we present a $\chi^2$-analysis in Section~\ref{dis:chi-square}.

The median random error is approximately 1~\% of the source flux density in band C and D, 2~\% of the flux density in band B and 3~\% of the flux density in band A. 
Thus the 3~\% systematic uncertainty typically make up the largest part of the error bars.

Additionally, the reference flux densities for the flux density calibrators have an uncertainty of approximately 3~\% (see Section \ref{dis:calibrators}). Since this introduces a constant bias that affects all sources and observations equally, it has not been included in the error bars.

\section{Flux density catalog}
\label{sec:results}

We obtained light curves for a large number of sources. 
Our catalog comprises 361 AGN, and of these 270 have data from at least five epochs. 
Data for these 270 sources are presented in Appendix~\ref{app:sourceplots}, Figs.~\ref{fig:all_lightcurves1}-\ref{fig:all_lightcurves5}.  
Flux density tables for all sources are available at the CDS, as well as on GitHub\footnote{\url{https://github.com/alvakinman/OSO_flux_monitoring}} where they will be continuously updated. 
Tab.~\ref{tab:maindata} shows an example of how the data are structured.

\begin{figure*}
    \centering
    \includegraphics[width=0.8\linewidth]{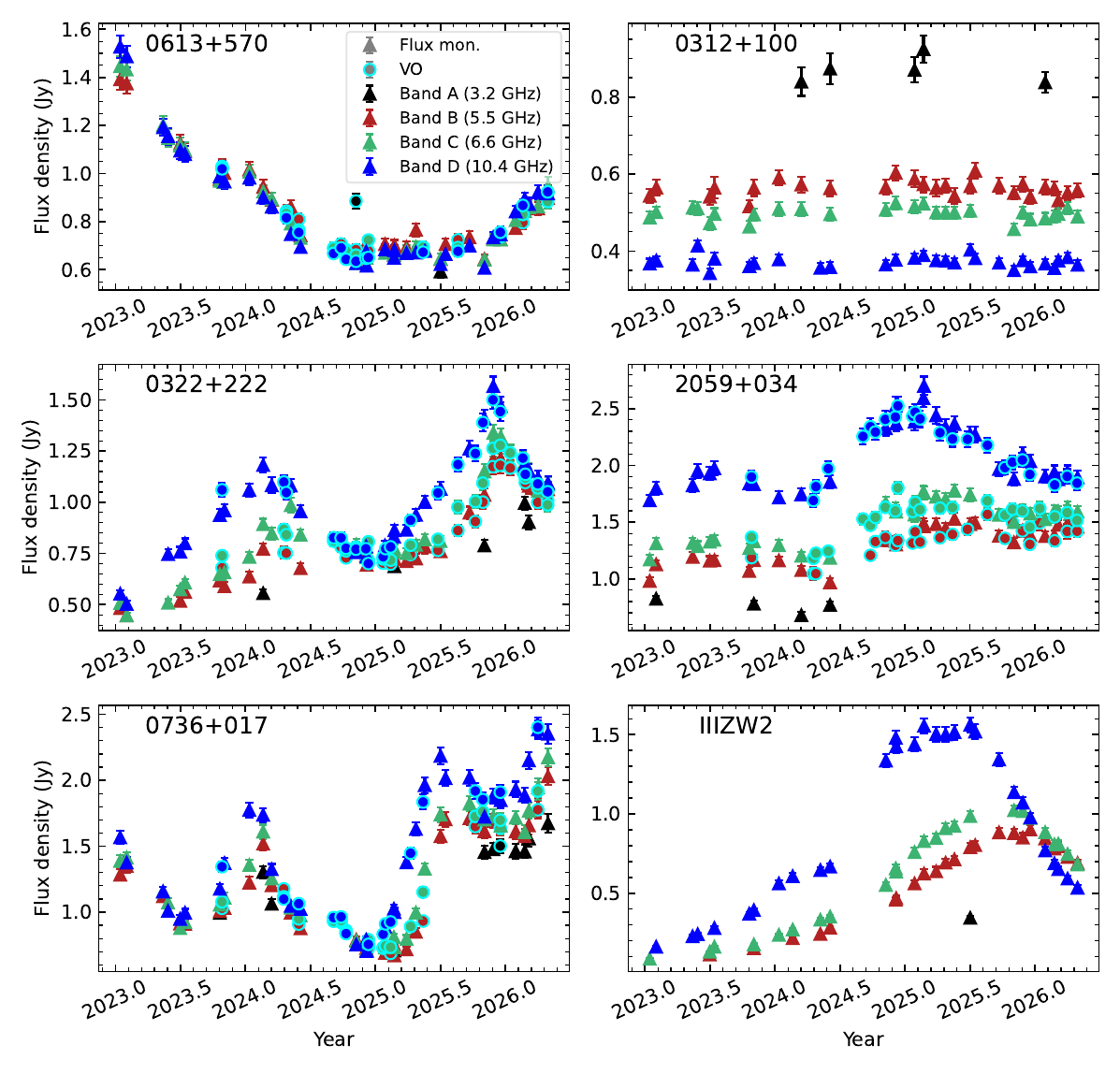}
    \caption{Example light curves from our flux monitoring experiments. Triangles represent results from dedicated FM experiments and circles from VO experiments. The four colors represent VGOS band A (3.2~GHz), B (5.5~GHz), C (6.6~GHz), and D (10.4~GHz).
The flux density of 0312+100 (top right) is relatively stable over the measurement period, while the other shown sources are more
variable. Note the differences in spectral index, both between sources (compare 0613+570, 0312+100, and 2059+034) and between
observations of the same source (compare 0322+222 in early 2024 and early 2025). }
    \label{fig:lightcurves}
\end{figure*}

A few representative light curves are shown in Fig.~\ref{fig:lightcurves}. 
The error bars are computed as described in Section~\ref{met:errors}, i.e. by taking the random error and adding 3~\% of the flux density in quadrature. The panels in Fig.~\ref{fig:lightcurves} exemplify different characteristics of the light curves in our catalog. Some sources are stable or slowly variable over time, such as 0312+100. Others undergo flaring events, such as 0322+222, 0736+017, and IIIZW2. 0613+570 has approximately equal flux density in all bands, while 0312+100 is brightest at low frequencies. 2059+034 is instead brightest at high frequencies.

\subsection{Radio source properties}

\label{res:properties}

\begin{figure}
    \centering
    \includegraphics[width=0.75\linewidth]{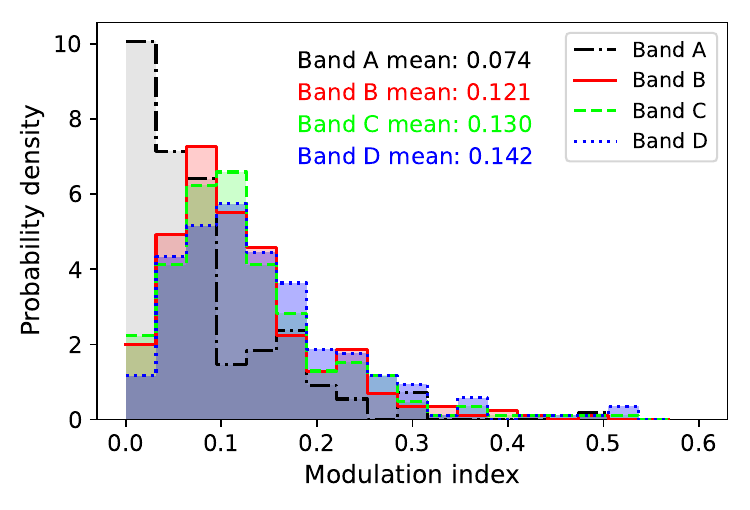}
    \caption{Distribution of modulation indices in each band. Histograms have been normalized to have an area of 1.}
    \label{fig:mod_index}
\end{figure}

The average flux densities of the sources in the catalog range from 0.09 to 75~Jy, but the vast majority of sources have a mean flux density below 5~Jy in all bands.
Most sources are significantly variable. 
We quantify the variability of the sources using the modulation index,
\begin{equation}
    \mu=\sigma_F / \Bar{F_\nu},
\end{equation}
i.e. the standard deviation of the flux density divided by the mean flux density \citep[see e.g.][]{shabala2014}. 
We find that 64~\% of sources have a modulation index above 0.1 in band~D, while 20~\% have a modulation index above 0.2. 
The distribution of modulation indices is presented in Fig.~\ref{fig:mod_index}. 
There is a tendency towards higher variability in higher frequency bands, with the largest average modulation indices found in band D. 
It is also useful to quantify the difference between highest and lowest flux density during the measurement period, as a fraction of the mean. 
There we find that 87~\% of sources vary by at least 20~\% in band D during the measurement period, while 39~\% of sources vary by at least 50~\%. 
The source flux density can thus change significantly within a few years. Note that only sources with at least five flux density measurements were included in computing these numbers. 
The most extreme example is the source {IIIZW2} (see Fig.~\ref{fig:lightcurves}), which increased in flux density by an order of magnitude between January 2023 and January 2025.

\begin{figure}
    \centering
    \includegraphics[width=0.85\linewidth]{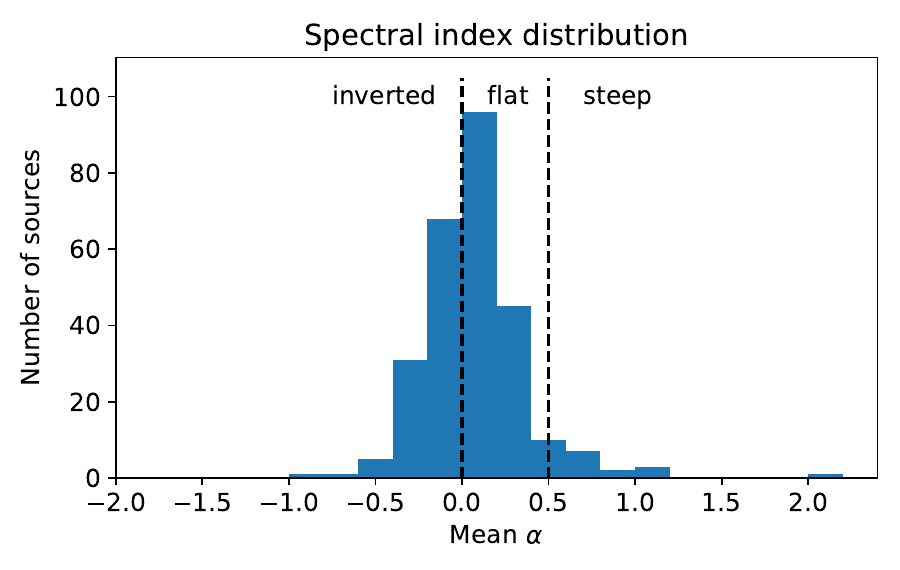}
    \caption{Histogram of mean spectral indices. Most sources have spectral indices around 0. The dashed lines mark the borders between inverted, flat and steep spectral index.}
    \label{fig:alpha_hist}
\end{figure}

The sources also show a variety of spectra, which can change over time. ~
Since the radiation from these sources is dominated by synchrotron radiation, it follows a smooth spectrum that for most sources can be described as a simple power law:
\begin{equation}
    F_\nu \propto \nu ^{-\alpha}.
\end{equation}
The spectral index $\alpha$ varies significantly both between sources and over time for the same source. Sources with $\alpha<0.5$ are commonly defined as flat-spectrum  \citep[e.g.][]{healey2007, padovani2017}, while sources with $\alpha>0.5$ are defined as having a steep spectrum. Sources with $\alpha<0$ are classified as inverted. The majority of sources in this work are either flat or inverted, as seen in Fig.~\ref{fig:alpha_hist}. The source spectra have been fitted with a power law at each time step, after which the resulting $\alpha$ values have been averaged. 
Only 6~\% of the sources have  a steep spectrum. 
This is expected given that a flat or inverted spectrum indicates synchrotron self-absorption in an optically thick plasma. 
The defining sources of ICRF3 were selected partly based on their compact morphology \citep{charlot2020}. This means that the emission mainly originates from the compact parts of the source, which are likely optically thick. 

\begin{figure}
    \centering
    \includegraphics[width=0.92\linewidth]{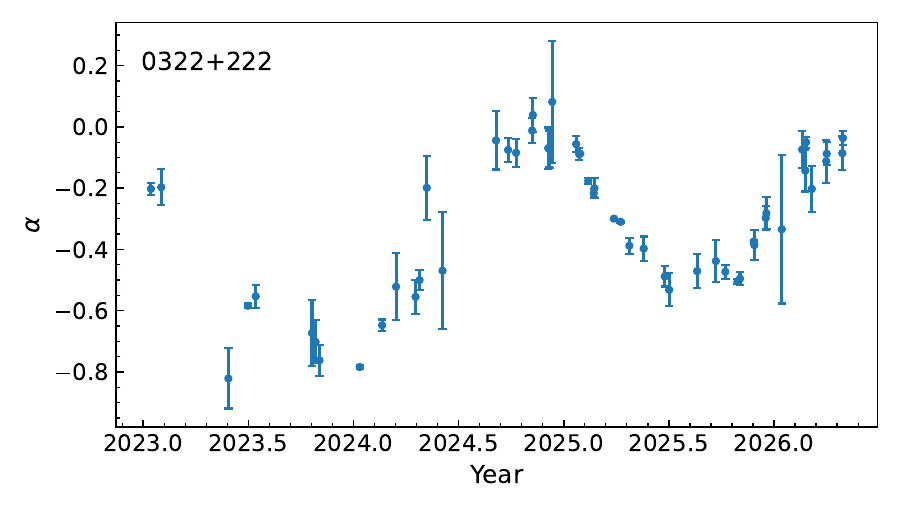}
    \caption{Spectral index for 0322+222 computed by fitting a power law to the flux density measurements at each epoch. The two dips correspond to periods of increasing flux density as seen in Fig.~\ref{fig:lightcurves}.}
    \label{fig:alpha_example}
\end{figure}

An example of spectral index variation over time is presented in Fig.~\ref{fig:alpha_example}, showing the spectral index of 0322+222. 
The spectral index varies smoothly between flat and inverted, with a lowest value around $-0.8$. 
Two prominent dips are seen centered at times 2024.0 and 2025.6, corresponding to periods of increasing flux density (see Fig.~\ref{fig:lightcurves}). In Fig.~\ref{fig:lightcurves} we can further note that the spectrum of both 0736+017 and {IIIZW2} switches between flat and inverted as well.

%- - - - - - - - - - - - - - - - - - - - - - - - - - - - - 
\subsection{Motivation for absolute flux density scaling}
%- - - - - - - - - - - - - - - - - - - - - - - - - - - - - 
Flux density scaling using  radio sources is typically not done in VLBI, mainly due to the lack of calibrators that are stable on small angular scales \citep[e.g.][]{janssen2019}. 
System temperature calibration is used as a replacement. 
In this work, both system temperatures and absolute flux density scaling have been used. 
The benefits of the absolute scaling are illustrated by examining the unscaled data. 
Before scaling, the data show systematic flux density variation both between sub-bands and between experiments.

Firstly, imperfect calibration of the noise diode can cause a bias of the system temperature in each sub-band. 
This bias is removed by introducing scale factors, as illustrated in \citet[][Fig.~1]{kinman2025}.

Secondly, there are amplitude variations between experiments in the unscaled data. 
This can be most clearly seen by studying the unscaled flux densities of the flux calibrators. 
The light curves of different radio sources are expected to be independent of each other, meaning that any correlation between them is most likely induced by the instruments or analysis. 
We find a high degree of correlation (Pearson correlation coefficient above 0.9) between the unscaled flux densities of different calibrator sources.
After applying the scale factors, the correlation is reduced.
This is illustrated in Appendix~\ref{app:correlation}, Fig.~\ref{fig:correlation_bigplot}.

The scale factor correction is usually on the order of 10~\%.
The mean scale factor for each sub-band is typically in the range 0.9-1.1, with a standard deviation of 0.05--0.10 in band B--D. Band~A scale factors are more variable.

%- - - - - - - - - - - - - - - - - - - - - - - - - - - - - - - - - - 
\section{Estimating signal-to-noise ratios of geodetic experiments}
%- - - - - - - - - - - - - - - - - - - - - - - - - - - - - - - - - - 
\label{sec:snr}
A major motivation for this work is to improve scheduling of geodetic VGOS experiments. 
With accurate and updated flux density measurements, the S/N of a scheduled observation can be calculated more accurately and the observing time can be optimized. 
To test whether our flux densities are useful for predicting S/N, we computed theoretical S/Ns for previously observed VGOS experiments and compared the results to the measured S/Ns.

\subsection{Analysis}

The expected S/N of an experiment can be calculated by 
\begin{equation}
\label{eq:snr}
    S/N = \eta F_{\nu} \frac{\sqrt{2\Delta \nu t}}{\sqrt{\rm SEFD_1\cdot SEFD_2}},
\end{equation}
which is simply a rearrangement of Eq.~\eqref{eq:obstime}. 
The efficiency factor $\eta$ is a product of several contributions: $\eta = \eta_{\rm Q} \eta_{\rm C} \eta_{\rm B}$.
$\eta_{\rm Q}=0.881$ is the quantization efficiency for two-bit sampling \citep{thompson2017}, $\eta_{\rm C}=0.97$ is an estimated correlation efficiency for DiFX \citep{sked-manual}
and $\eta_{\rm B}$ is the bandpass efficiency.
Since the bandpass response is lower towards the edges of the sub-band, the S/N is lower than for an ideal rectangular bandpass \citep{mccallum2022}.
The OTT, as well as the majority of the other VGOS stations, have DBBC3 back ends. We assume that the DBBC3 unit belonging to ONSA13NE is representative, and estimate the bandpass efficiency $\eta_{\rm B}$ to 0.92 (see Appendix \ref{app:bandpass}). The final efficiency factor becomes $\eta = 0.786$. However, we note that some stations instead use RDBE back ends, which are known to have a more rectangular bandpass than the DBBC3. Thus, the true $\eta_{\rm B}$ factor may be larger for baselines involving such stations.
Furthermore, we used $\Delta\nu = 2\cdot8\cdot 32$~MHz, to account for the two polarizations and eight sub-bands in each band.

We compared predicted to observed S/N for all VO sessions observed in 2025. 
Predictions were computed both based on the flux density values from this work and based on the currently used scheduling flux density catalog \texttt{flux.cat.vgos}, version 2023-01-01 \citep{schartner2025}\footnote{\label{vieschedpp}\url{github.com/TUW-VieVS/VieSchedpp_AUTO/blob/master/CATALOGS_VieSchedpp}}.

The flux density catalogs used in geodetic scheduling by the softwares sked and VieSched++ can contain two types of flux density data.
The first type is Gaussian models (labeled M) that treat the source as an elliptical Gaussian with a peak value, full width half maximum along two axes and orientation angle. 
The second type is binned models (labeled B) that contain one flux density value per bin in projected baseline length. For example, every observation of the source with a projected baseline between 0 and 1000~km is assigned one flux density value, while observations with a projected baseline between 1000 and 2000~km are assigned another value. 
These models do not account for baseline orientation. 
The catalog \texttt{flux.cat.vgos} only contains B type models, with 1000~km wide bins ranging from 0 to 13000~km.

From the monitoring campaign described in this work, we only gain information about total flux densities. 
However, on longer baselines the sources will be resolved and less of the extended emission will be detected. 
We therefore expect the flux density that is visible to the interferometer (and therefore influences the S/N) to decrease when baseline length increases.

To get flux densities on longer baselines, we used the data in the catalog \texttt{flux.cat.vgos}. 
We scaled the flux density in the smallest baseline bin, i.e. for projected baselines between 0 and 1000~km, to match the flux density value from our measurement. 
The flux density values in longer baseline bins were then scaled by the same factor. 
This assumes that a change in flux density has the same relative size on all spatial scales. 
The assumption is likely most accurate for very compact sources.
We expect discrepancies for other sources, but still consider it a useful approximation.

For simulating geodetic scheduling, the flux density values used must be the values known at the time of the scheduling. 
In this analysis, the flux density catalog used to compute predicted S/Ns is derived from the latest flux experiment that was observed at least 30 days prior to the VO session. 
If the flux density values for a source are missing in one or two bands, the missing values are approximated by fitting  a power-law spectrum to the existing values. 
If three or four flux values are missing, the values are instead taken from the catalog for the previous experiment.

Information about observed S/N, observing time, station coordinates and source coordinates for the VO sessions were obtained from VGOS database files, available from CDDIS\footnote{\url{https://cddis.nasa.gov/archive/vlbi/ivsdata/vgosdb/}}. 
Since the VGOS databases only contain total S/N, the band-wise S/N was first computed according to Eq.~(2) in \cite{corey2022},
\begin{equation}
    (S/N)_{\rm band} = (S/N)_{\rm tot} \left |\frac{\sqrt{M}}{\sum_{m=1}^M V_m} \right | \left | \frac{\sum_{n=1}^N V_n}{\sqrt{N}}\right |,
\end{equation}
where $M$ is the total number of sub-bands, $N$ is the number of sub-bands in the given band, and $V_i$ is the complex visibility of each sub-band.

We obtained SEFD values from the catalog \texttt{equip.cat.vgos}, currently used in VGOS scheduling\footnote{\label{vieschedpp}\url{github.com/TUW-VieVS/VieSchedpp_AUTO/blob/master/CATALOGS_VieSchedpp}}.
The SEFD value for each band is given as a function of elevation $\epsilon$, with four station-and band-dependent parameters $S, c_0, c_1, y$:
\begin{equation}
\label{eq:sefd}
    {\rm SEFD(\epsilon)} = S\cdot\left (c_0 + \frac{c_1}{(\sin{\epsilon})^y} \right ).
\end{equation}
$S$ represents the SEFD at zenith, $c_0$ is the fraction of the zenith SEFD that is elevation-independent, and $c_1$ and $y$ describe the elevation-dependent SEFD variation \citep[][Ch.~5]{sked-manual}.

In summary, theoretical S/N values for each observation were computed as follows. 
Station coordinates, source coordinates and time from the VGOS database file were used to compute source elevation and projected baseline length. 
The elevation was used together with SEFD data from \texttt{equip.cat.vgos} and Eq.~\eqref{eq:sefd} to compute the SEFD of each antenna. 
The projected baseline length was used together with the flux density catalog (either \texttt{flux.cat.vgos} or a catalog based on measured values) to select the correct flux density for the source. 
Finally, the flux density $F_\nu$ and the SEFD values were used as input to Eq.~\eqref{eq:snr}. 
To be considered in this analysis, a source needed to have both an entry in \texttt{flux.cat.vgos} and flux density data from this work.

\subsection{Results}

To evaluate the quality of S/N predictions, the ratio of predicted S/N to observed S/N was computed. 
This ratio was then compared between Case~1, where the standard flux density catalog (\texttt{flux.cat.vgos}) was used in the prediction, and Case~2, where incrementally updated flux density catalogs were used. 

\begin{figure*}
    \centering
    \includegraphics[width=0.9\linewidth]{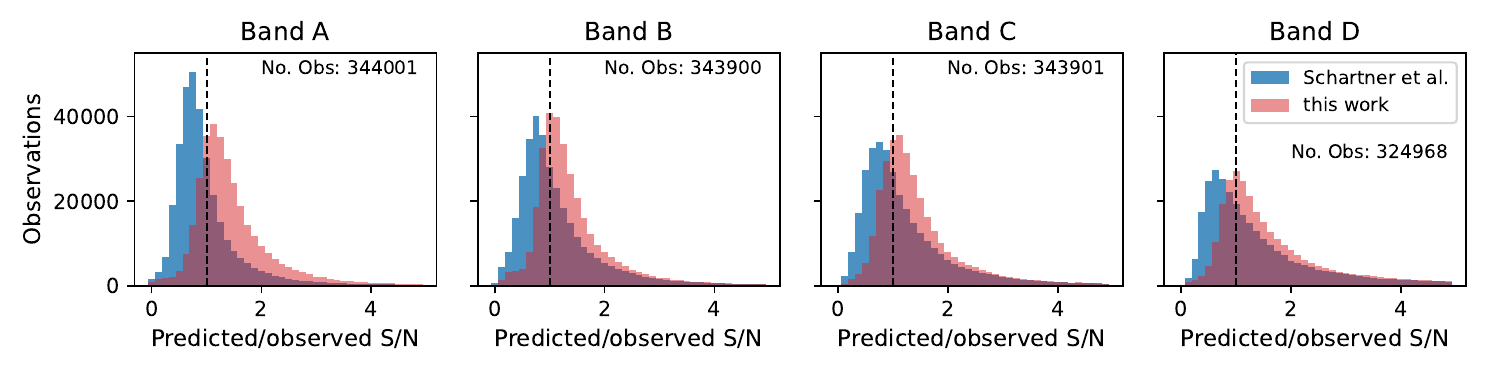}
    \caption{ Ratios of observed and predicted S/N for two cases: with flux densities from the constant VGOS catalog (in blue) and with flux densities from this work (in red). The dashed line represents a ratio of 1. All observations from VO experiments in 2025 are included. The flux densities from this work lead to higher estimated S/Ns than the standard VGOS catalog, but the distribution peak is close to 1.}
    \label{fig:snrhist_all}
\end{figure*}

Fig.~\ref{fig:snrhist_all} shows the distribution of S/N ratios in each band, for both Case~1 and Case~2. ~
The dashed line marks a ratio of unity, which would be achieved if the prediction was perfect. 
We find that the peak of the Case 2 distribution is close to unity, meaning that the flux densities from this work produce accurate S/Ns. The predicted S/Ns of Case~1 are seen to be systematically lower than Case~2.

\begin{figure*}
    \centering
    \includegraphics[width=0.9\linewidth]{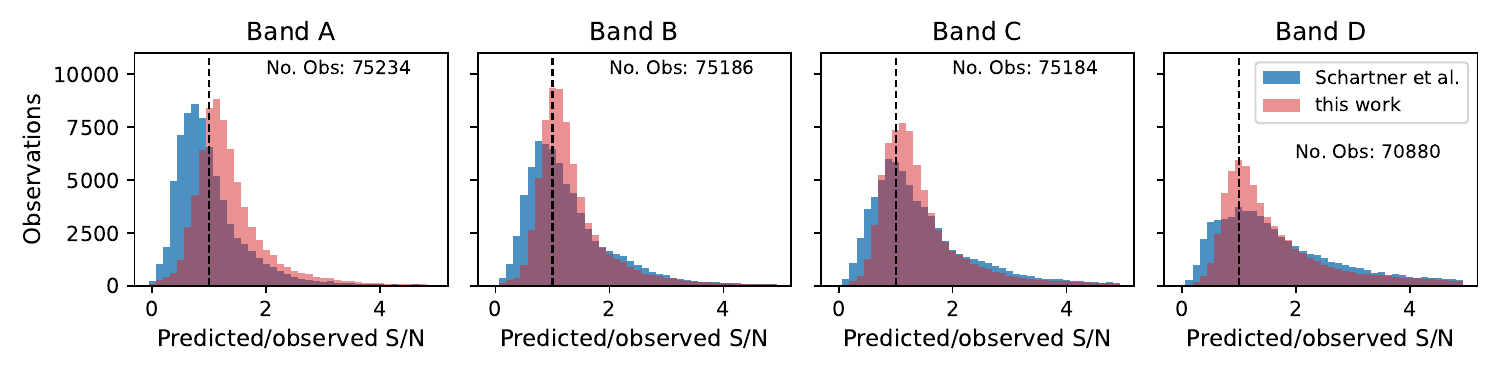}
    \caption{ Analogous to Figure~\ref{fig:snrhist_all}, but only the 20 sources with the largest flux density variability are included. While the predicted S/Ns based on this work are still higher on average, the dispersion is clearly smaller in band~B-D.}
    \label{fig:snrhist_20}
\end{figure*}

\begin{figure*}
\begin{subfigure}{\linewidth}
    \centering
    \includegraphics[width=0.9\linewidth]{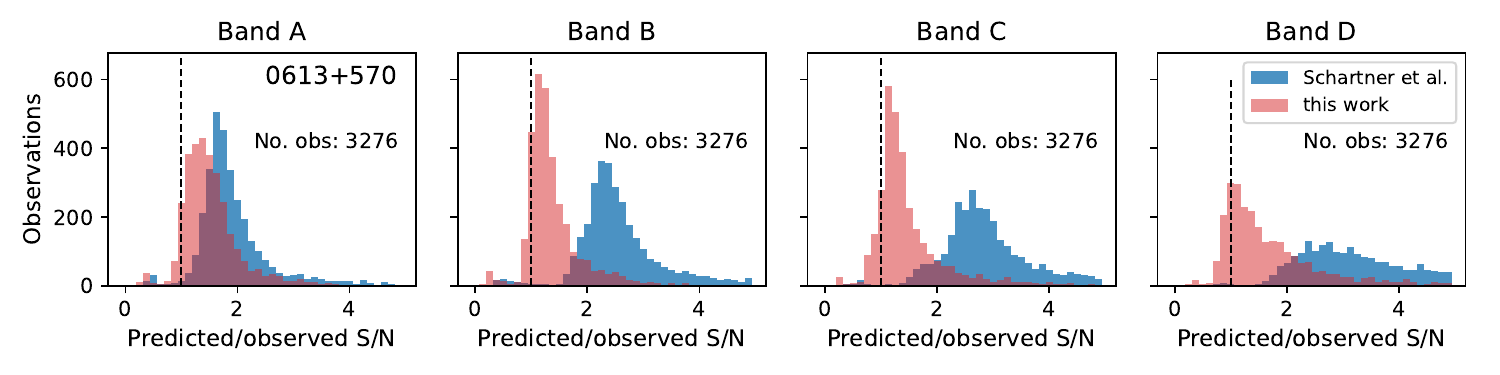}
    \label{fig:0613+570}
\end{subfigure}
\begin{subfigure}{\linewidth}
    \centering
    \includegraphics[width=0.9\linewidth]{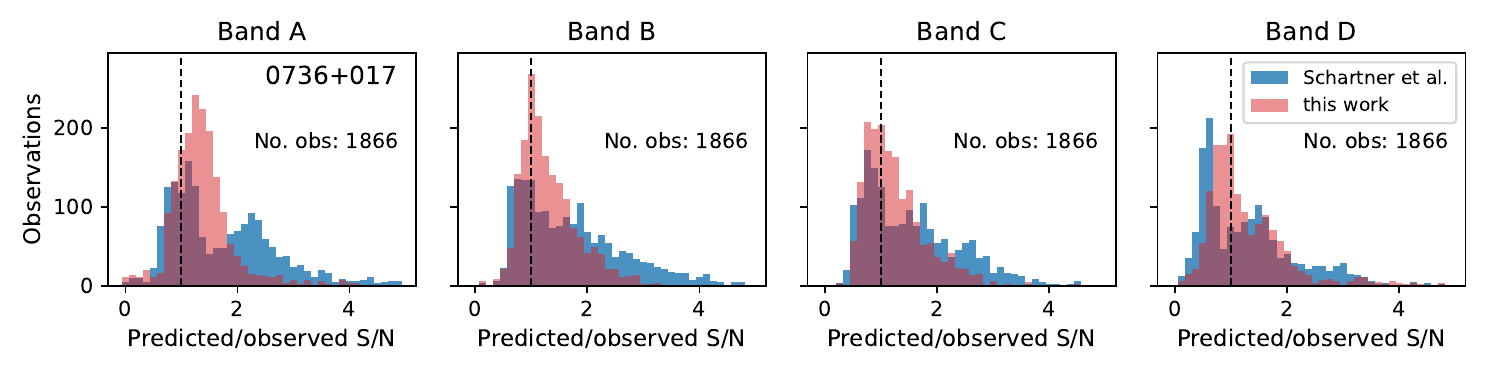}
    \label{fig:0736+017}
\end{subfigure}
    \caption{Analogous to Fig.~\ref{fig:snrhist_all}, but only including observations of 0613+570 (top) and 0736+017 (bottom). The S/N of 0613+470 is severely overestimated by using the standard catalog, while the catalog from this work comes closer to observed S/Ns. The standard catalog also overestimates the S/N of 0736+017 for a portion of the observations, while underestimating the S/N for others. The catalog of flux densities from this work gives more consistent S/N ratios.}
    \label{fig:snr_individual}
\end{figure*}

\begin{table*}[]
    \small
    \centering
        \caption{ Median and median absolute deviation (MAD) for the ratio of predicted and observed S/N.}
    \begin{tabular}{llllll|llllll}
    \hline\hline
        Selection & Band & \multicolumn{2}{c}{Median} & \multicolumn{2}{c|}{MAD} & Source & Band & \multicolumn{2}{c}{Median} & \multicolumn{2}{c}{MAD} \\
         &  &  Case~1 & Case~2 & Case~1 & Case~2  & & &  Case~1 & Case~2 & Case~1 & Case~2 \\
        \hline
        All obs. & A & 0.84 & 1.33& 0.26 & 0.36  &       0613+570  & A & 1.79 & 1.44 & 0.26 & 0.27\\
         &  B & 0.96 & 1.26& 0.36 &0.34          &                 & B & 2.46 & 1.26 & 0.36 & 0.20 \\
         &  C & 1.02 & 1.29& 0.42 &0.39          &                 & C & 2.86 & 1.30 & 0.52 & 0.24\\
         &  D & 1.19 & 1.49 &0.61 &0.61          &                 & D & 3.92 & 1.66 & 1.49 & 0.63 \\

         \hline
       All obs.,  & A & 1.00 & 1.00 &0.31& 0.27  &       0736+017  & A & 1.67 & 1.32 & 0.72 & 0.28\\
        median-scaled  & B & 1.00 & 1.00 & 0.38& 0.27  &           & B & 1.61 & 1.21 & 0.69 & 0.31 \\
         &  C & 1.00 & 1.00  &0.42& 0.30 &                         & C & 1.47 & 1.16 & 0.66 & 0.36\\
         &  D & 1.00 & 1.00 & 0.51& 0.41 &                         & D & 1.40 & 1.25 & 0.79 & 0.49\\
         \hline
        20 most var.  & A & 0.90 & 1.25 & 0.33& 0.33 \\
          sources & B & 1.14 & 1.21 & 0.43 & 0.31 \\
           & C & 1.28 & 1.28 & 0.52 & 0.38\\
          & D & 1.67 & 1.47 & 0.87 & 0.59 \\
          \cline{1-6}
    \end{tabular}
    \label{tab:snrratio}
\end{table*}

In Tab.~\ref{tab:snrratio}, the median and median absolute deviation (MAD) of each case is shown. ~
Note that the mean and standard deviation are not useful metrics for this distribution, since the data contain large outliers. 
A small part of the observations fails, e.g. due to technical issues, producing observed S/N values much smaller than theoretically expected. 
These form a tail of the distribution that reaches above 10. 
To avoid that these comparatively few points disproportionately affect the average and dispersion measures, the median-based measures are used.
From the first part of Tab.~\ref{tab:snrratio} it is clear that the median for Case 2 is higher than for Case 1, and that the dispersion of ratios is similar (except for band A, where Case~1 has a smaller dispersion).

However, the relative dispersion is notably lower with the new flux density catalogs.
If both distributions are divided by their respective medians, the distribution based on the new flux density catalogs is narrower. 
The MAD for Case~2 is up to 29~\% lower than Case~1 when median-scaling is applied, as can be obtained from Tab.~\ref{tab:snrratio}.

The advantage of Case 2 emerges further when considering only the most variable sources.
Fig.~\ref{fig:snrhist_20} contains the same S/N ratios as Fig.~\ref{fig:snrhist_all}, but the observations are limited to the 20 sources that have the largest modulation index (see Sect.~\ref{res:properties}) in band~D. 
Here it is evident that the Case~2 histograms have a smaller spread in band B, C, and D. 
The MAD values listed in Tab.~\ref{tab:snrratio} are clearly smaller as well.
This is without the median scaling, so if the constant bias between the cases was removed there would be an even larger advantage for Case~2.

Finally, we present similar histograms for two individual sources, to demonstrate the effect on the S/N predictions if the true flux density is very different from the fixed value from \cite{schartner2025}.
The top row of Fig.~\ref{fig:snr_individual} shows the ratio between predicted and observed S/N for the source 0613+570.
The source belongs to the group of the 20 most variable, with a modulation index of 0.26. 
In Fig.~\ref{fig:lightcurves}, it can be seen that the flux density of the source has steadily decreased since 2023, stabilizing around 0.6 Jy in 2025.
We find that the typical predicted S/N is several times higher than the observed S/N when using the \citeauthor{schartner2025} catalog. The flux density catalogs from this work give more accurate and precise S/N predictions.

The bottom row of Fig.~\ref{fig:snr_individual} shows the S/N ratio for the source 0736+017.
This source is also among the most variable, with a modulation index of 0.35. 
The light curve is also presented in Fig.~\ref{fig:lightcurves}. 
By using the standard catalog to predict S/N, we obtain a bimodal distribution of S/N ratios. 
The explanation is that the source has increased in flux density by roughly a factor of 2 during 2025, with the two peaks of the distribution corresponding to the early and late part of the year. 
With the flux density catalogs from this work, the flux density increase can be tracked and a more narrow S/N ratio distribution is obtained.

\section{Discussion}

\label{sec:discussion}
We presented a set of light curves at the four currently used VGOS bands, centered at 3.2, 5.5, 6.6, and 10.4~GHz, including 361 ICRF3 sources.
Here, we discuss errors and biases of the flux densities, 
followed by the implications of the results for geodetic scheduling, and finally the future development of our monitoring program.

\subsection{Systematic errors in flux density measurements}

\label{dis:errors}
There are a few different systematic error contributions in the flux density measurements. 
For example, there are two different error sources that are related to the elevation angle.

Firstly, a flat gain curve has been used for calculating gains. 
We know, however, that the OTT experience some elevation-dependent gravitational deformation \citep{loesler2019}.
This should also result in gain that varies with elevation. 
Recently, gain curve measurements were performed, 
which can be used to refine the flat gain curve estimate for future experiments.

Secondly, the atmospheric opacity varies with elevation. 
This is most significant in VGOS band~D, where low elevation sources may lose a few percent of their flux density due to atmospheric attenuation. 
We plan to correct for this in future observations by using atmospheric opacity measurements of OSO's water vapor radiometers.

We note that the elevation-dependent effects are mitigated for many sources since they are observed in several scans with different elevations.
However, low-declination sources are only visible from OSO at low elevation.
The above systematic errors are thus expected to affect them more.

Additionally, the elevation-dependent effects could influence the scale factors, as the calibrator sources are observed at different elevations in different experiments. For FM experiments we expect this effect to be small since multiple spaced-out calibration scans are averaged.
The VO experiments could be more affected since each calibration source is only observed once.

Lastly, the most troubling error source is the presence of disturbing electromagnetic radiation. 
The two highest-frequency sub-bands in band~A are almost constantly affected, leading us to exclude them completely from the analysis.
The rest of VGOS band~A is also severely affected, causing this band to be flagged in the majority of our measurements.
Automatic flagging of RFI is imperfect, and inspecting all scans manually to flag unwanted signals is not practically feasible. 
One way to improve the situation would be to slightly change the spectral setup in the local sessions. 
Instead of observing the two highest band~A sub-bands, one could try to find two other frequency windows within band~A that are cleaner.

\subsection{Chi-square analysis}

\label{dis:chi-square}
To verify that the added systematic error of $3~\%$ is reasonable, we performed a reduced $\chi^2$ analysis on a few sources with stable flux densities. 
In this analysis, we assumed that the  true flux density of the source is constant, and that any variations are caused by measurement errors (random + systematic). 
The flux calibration sources have been included here, but note that they have been rescaled. 
For each calibrator source, we calculated a new scaling factor based only on the other two sources, such that the variation for the analyzed calibrator was not artificially reduced. 

The results of the analysis are shown in Tab.~\ref{tab:chi2-3}. 
In general the values are smaller than 1, indicating that the error is overestimated rather than underestimated. ~
For 0312+100 the values are very close to 1. 
The reduced $\chi^2$ value in band~A is missing for 0312+100 and 2111+400 since there were too few data points for an accurate estimate.

\begin{table}[]
    \centering
    \small
        \caption{Reduced $\chi^2$ values}
    \begin{tabular}{l|llll}
    \hline \hline
    Source & Band A & Band B & Band C & Band D \\
    \hline
       3C147  & 0.54 & 0.30& 0.29 & 0.45 \\
        3C286  & 0.46& 0.46& 0.44  & 0.66 \\
         3C295  & 0.36& 0.31& 0.30 &  0.54 \\
         0312+100  & -& 1.00 & 0.97 & 1.02 \\
          0552+398  & 2.87 & 0.98 &0.38 & 0.74  \\
           2111+400  & -& 0.65 & 0.43 & 0.66 \\
           \hline
    \end{tabular}
    \label{tab:chi2-3}
\end{table}

\subsection{Potential biases due to scan averaging}

\label{dis:averaging}
In this work, scans were averaged by amplitude (as in \cite{varenius2022} and \cite{kinman2025}) and not by visibility. 
The aspects that factored into this decision will be described below.

In order to average scan visibilities, it is important that the phases are correctly calibrated. 
If there are any systematic phase differences between scans, the visibilities will be averaged incoherently and the amplitude will be underestimated. 

In this work, phase calibration was done by performing a multi-band fringe fit on a few bright scans only.
This removes the main linear phase trend caused by clock rate errors. 
The atmosphere-dependent phase variations are already negligible due to the short distance between the OTT. 
There is, however, a third effect that makes the phase change over time in these experiments. 
In the majority of the experiments, a phase oscillation with a period of two hours can be seen. 
It is most likely caused by periodic temperature variations in the control room at OSO, which impacts the DBBC3 units that digitize the signal. 

We made attempts to fit a sinusoidal function to the calibrated phases, but this did not yield sufficiently accurate results. 
A second way to deal with the issue is to perform a multi-band fringe fit on every scan. 
This comes with its own drawbacks, as it prevents us from observing sources that are too faint for fringe-fitting. 
Additionally, there is a risk that the fringe fit algorithm converges to the wrong solution when used on moderately faint sources. 
This was found to affect some low-elevation sources in particular. 

Given this, we chose not to attempt to remove the phase oscillation. 
Instead we averaged the visibilities within a scan and then averaged the scan amplitudes. 
To achieve coherent visibility averaging within a scan, it is sufficient that the phase is stable for the duration of the scan, which is on the order of seconds to minutes. 
This is fulfilled even in the presence of the phase oscillation. 

The drawback of averaging amplitudes is that the amplitude of low-S/N scans may be overestimated. 
While the real and imaginary part of the visibility each follow a Gaussian distribution, the amplitude instead follows a Rice distribution \citep[][Chapter 9.3.3]{thompson2017}. 
In general, the mean of the Rice distribution is higher than the true (noise-free) visibility amplitude. 
However, when the S/N is high, the Rice distribution approaches a Gaussian. 
In other words, we will obtain the correct visibility amplitude by averaging individual scan amplitudes, as long as the S/N is high enough.

What is then a sufficient S/N for the amplitude-averaging method to be valid? 
By Eq. (9.65) in \citep{thompson2017}, we find that an $S/N<7$ leads to an amplitude error of $> 1$~\%. 
Assuming a typical SEFD for the OTT of 2000~Jy, an S/N of 7 corresponds to a 0.41~Jy source observed for 30 seconds or a 0.20~Jy source observed for 120 seconds. 
In the VO experiments, there are six sources that at some point in time are fainter than 0.41~Jy. 
The FM experiments have 28 sources that at some point were below 0.20~Jy, but the vast majority of the observations fulfill the condition. ~
However, the amplitude overestimation has a major impact on the cross-polarization amplitudes, as the true amplitudes are expected to be on the same order as the noise. 
The cross-polar amplitudes are greatly overestimated for the VO experiments in particular, since the short scan durations give a higher noise per scan than in the FM experiments. 
This explains why the previously used cross-polar outlier criterion flagged so much data in the VO experiments in particular (see Appendix~\ref{app:outlier}). 
With the new deviation-based outlier criterion, the cross-polar amplitudes no longer impact the flux density values. 
It is thus of no importance whether they are estimated correctly.

\subsection{Stability and accuracy of the calibrator flux densities}

\label{dis:calibrators}
The flux densities of the calibrator sources are taken from \cite{perley2017}. 
We assume that the flux densities of 3C147, 3C286, and 3C295 have not changed significantly since its publication. 
\citeauthor{perley2017} report that 3C286 and 3C295 are known to be stable over long timescales, while 3C147 is known to be slowly variable. 
However, the authors also present a comparison of their flux density values to the flux density scale of \cite{baars1977}. 
Even though these works are separated in time by 40 years, the reported flux densities for 3C286, 3C295, and 3C147 differ by less than 5~\% from their 2017 values in the frequencies relevant for VGOS.  
Furthermore, recent work presents evidence of the continued stability of 3C286, although at higher frequencies \citep{kam2025}. 
We therefore argue that the flux scale of \citeauthor{perley2017} can still be used, until a more recent version is available. 
If one of the calibrators started to vary in flux density, it would be detectable in the obtained scaling factors for that source. 
So far, the scaling factors obtained from the three calibrators agree well, indicating that the flux densities have stayed close to their values reported in \cite{perley2017}. 
The uncertainty of the flux density scale is reported to be around 3~\% in the VGOS frequency range.

\subsection{Spectral index computation}

We used a simple power-law model to describe the spectral index. 
While this works well for the majority of our sources, there are a few sources, for example 0552+398, that show evidence of a spectral peak within the monitored frequency range. 
To model the spectrum of such sources, a peaked synchrotron self-absorption (SSA) spectrum is more appropriate,
\begin{equation}
    S_\nu = C \left( \frac{\nu}{\nu_p}\right )^\alpha \left ( \frac{1-e^{-\tau}}{\tau}\right ) ,
\end{equation}
where $\nu_p$ is the frequency where the source becomes optically thick, $\tau = (\nu/\nu_p)^{-(\beta+4)/2}$ is the optical depth and $\alpha = -(\beta-1)/2$ is the spectral index. 
The parameter $\beta$ is the power law index of the electron energy distribution  \citep{tingay2003,shao2022}. 
However, this spectrum has three free parameters: $C$, $\beta$ and $\nu_p$. 
In the best case we have access to four flux density measurements per source and experiment, and in the typical case we have only three. 
Fitting this spectrum thus gives less robust results compared to the simple power law, which has only two parameters. 
For the simple analysis presented in this paper, e.g. estimating the fraction of steep- and flat-spectrum sources in our sample, we have thus elected to use the simple power law. 
For investigating the spectral properties of individual sources, however, the SSA spectrum or another peaked model \citep[see ][]{shao2022} may be needed.

\subsection{Signal-to-noise-ratio estimation and geodetic scheduling}

We found that the flux densities in this work can produce accurate S/N predictions.
When considering the ratios between predicted and observed S/N, the peak of the distribution is close to one.
The median is, however, significantly larger than one. This means that many observations perform worse than expected, which may for example be caused by RFI signals increasing the SEFD of a station.
The S/N requirements in scheduling are usually set to be greater than needed to account for such variations.

The median S/N ratios, and by extension the flux densities, of \cite{schartner2025} are lower than the flux densities derived in this work, by a factor 0.62, 0.77, 0.79, and 0.80 in band A, B, C, and D respectively.
It is important to note that \citeauthor{schartner2025} did not directly measure flux densities. Rather, they used observed S/N values from geodetic VLBI sessions and Eq. \eqref{eq:snr} to derive flux density estimates. 

Part of the flux density difference can be explained by the choice of the efficiency $\eta$ in Eq. \eqref{eq:snr}. 
\citeauthor{schartner2025} do not include the bandpass efficiency in their $\eta$ value.
This affects the computed flux densities by 8~\%.
When deriving a flux density catalog for scheduling by the method of \citeauthor{schartner2025}, it is most important that the $\eta$ value used matches the $\eta$ used in the scheduling software. 
The bandpass efficiency is not included by default in the geodetic softwares sked and VieSched++, and thus it is warranted to exclude it in the flux density derivation as well.
Our flux densities on the other hand are independent of $\eta$, and therefore the S/N needs to be computed with an $\eta$ that is true to the real performance of the VGOS system.

The choice of $\eta$ cannot explain all of the difference between our flux densities and \cite{schartner2025}. There are a few other aspects to consider.

It is not likely that the S/N bias reflects an error in our total flux densities. 
The observations of calibrator sources ensure that the normalization of the flux densities is accurate.
However, in our S/N analysis we equated the flux density on baselines up to 1000~km with the total flux density of the source. 
The angular resolution achieved with a 1000~km baseline is 7~mas in band~D, while the OTT baseline is sensitive to angular sizes of 100~as.
If a source has extended components larger than the VLBI beam, they will contribute to the total flux density measured by the OTT but not to the S/N on a longer baseline. 
It is possible that this causes overestimation of the long-baseline flux densities for some sources.

Furthermore, SEFD values used for deriving flux densities can have an impact on the results. \cite{schartner2025} derived SEFD values from measured system temperatures, which may be imperfectly calibrated and have biases between stations. In our flux density catalog, we were able to correct for such effects using flux density calibrator sources.

As we can see from figures~\ref{fig:snrhist_all}--\ref{fig:snrhist_20}, the spread in predicted/observed S/N ratios is significant even when the flux density values are improved. 
A major part of the explanation can probably be found in the treatment of longer baselines. 
The long-baseline model used here is outdated, as it was derived from data observed in 2022. Furthermore, it does not take baseline orientation into account. 
More accurate source structure modeling is needed to get realistic flux densities on all baselines.

One avenue to get structure models is to use the method in this work to evaluate the correlation amplitudes on longer baselines. 
There is, however, some development needed to make the method applicable in the presence of source structure effects and without available absolute calibrators. 
A more straight-forward option would be to use the method of \cite{schartner2025} to update the flux density catalogs with regular intervals \citep[where work is ongoing by][]{chakraborty2025}, with total flux density measurements used to track variations in between these updates.

A third option for updating the source structure models is to image the sources using the VGOS network. 
Such work is already in progress, see e.g. \cite{xu2021, schartner2023, perez-diez2024}. 
As most VGOS stations now measure system temperatures, imaging could also provide total flux density. 
Still, flux density monitoring with the OTT could be a complement to these experiments, as it requires fewer stations simultaneously observing, and uses less correlation resources. 
It also requires less observation time per source, as there is no need to achieve good $uv$-coverage.  
It could thus be done more often than full imaging experiments. 

Finally, source models alone cannot guarantee good S/N predictions. 
Regularly updated SEFD values for the stations are also needed. 
Since the SEFD is affected by weather and instrumental issues that cannot be known in advance, there is a limit to how accurate S/N predictions can become.

\subsection{Development of the monitoring program}

In this section, we consider how the monitoring sessions can best serve the needs of the geodesy and astronomy communities.
The flux monitoring experiments described in this paper primarily target ICRF3 defining sources. 
However, the regular VGOS sessions observe many sources that are not part of this set. 
To obtain a complete and current VGOS flux density catalog, it will be necessary to start monitoring a wider selection of sources. 
Although flux density values can be obtained from the VGOS sessions themselves,
the time from observation until correlated data are publicly available can be weeks to months. 
Additionally, not every source is observed in every experiment. 
The sources that are monitored only in VO sessions will thus not have the most current flux density values available.

For the reasons above, it would be beneficial to extend the source catalog of the monitoring program. 
However, ideally this should not happen at the expense of the already monitored sources, since long-term, consistent monitoring will increase the value of the data for astronomical use.

To be able to observe a larger number of sources, one option would be to involve more stations in the monitoring program. 
For example, the other VGOS twin telescope pairs could run independent flux monitoring experiments with different source catalogs. 
A few test observations involving the Wettzell and Ny-Ålesund VGOS antennas have been performed in 2025 \citep{kinman2025b}. 
Lastly, the source catalog monitored by the OTT is naturally limited to the northern sky. 
In order to monitor all sources relevant for VGOS, participating stations on the southern hemisphere would be needed.

Another issue is the ongoing effort to find new VGOS operational frequencies, to 1) make better use of the total bandwidth available to VGOS, and 2) minimize the disturbance from active users of the frequency bands \citep{hase2022,bernhart2025}. 
Once the new frequency setup has been determined and fully tested, the VO bands will shift and the flux density monitoring sessions should also cover the new bands to keep their relevance for VGOS scheduling. 
However, as mentioned previously there is also a value in obtaining long, consistent measurement series for astronomical purposes. Therefore it may be warranted to observe both the new and old VGOS frequency setup in the future.

\section{Conclusions}

\label{sec:conclusions}
We presented a flux density catalog for geodetically relevant AGN with simultaneous measurements in four radio frequency bands centered at 3.2, 5.5, 6.6, and 10.4 GHz. 
The measurements were made using the OTT, normally used for geodetic VLBI, as a single baseline interferometer. 
The flux densities in the catalog are obtained both from dedicated FM sessions and from VO experiments, where only the Onsala baseline has been analyzed.

The catalog contains 361 sources in total, of which 270 have data from at least five epochs. 
The majority of the sources have a flat or inverted radio spectrum. 
While all bands were observed for all sources, the data in band A are sparse due to unwanted radio signals around the observatory site.

The data from the monitoring program can be used to track  flux density variation over time, as well as spectral index variation. 
Many sources are significantly variable in the frequencies relevant for VGOS, in particular at 10.4~GHz. 
This underlines the importance to regularly update the flux densities used for geodetic scheduling. 

We used the flux densities of this work to compute theoretical S/N values for previous geodetic sessions. 
Our flux densities produce S/N estimates generally consistent with observed S/Ns, with a smaller relative spread than from the standard catalog used for VGOS scheduling.
The benefit of our results is most evident for the most variable sources, where the flux densities from this work produce a smaller absolute spread in S/N predictions.

There are still significant discrepancies between predicted and observed S/N in the VO experiments from 2025. 
Improved source models are one avenue to address this, but SEFD values also need to become more accurate.

In the future, we hope to include more VGOS stations in our flux density monitoring efforts. 
This would allow more sources to be covered, and potentially produce correlation amplitudes on longer baselines. 
Imaging and/or other methods to obtain regularly updated source structure models are also needed in order to more accurately predict signal-to-noise ratios in VGOS experiments.

The catalog presented in this paper will be continuously updated. 
Our aim is that it finds use both in geodesy and AGN astronomy.

\section*{Data availability}
%Tables xx and xx are only available in electronic form at the CDS via anonymous ftp to cdsarc.u-strasbg.fr (130.79.128.5) or via http://cdsweb.u-strasbg.fr/cgi-bin/qcat?J/A+A/.
The flux density data presented in this paper (Tab.~\ref{tab:maindata}), as well as Tab.~\ref{tab:name-translation}, are available in electronic form at the CDS via anonymous ftp to \url{cdsarc.u-strasbg.fr} (130.79.128.5) or via \url{http://cdsweb.u-strasbg.fr/cgi-bin/qcat?J/A+A/}.
%The full data presented in this paper is available at the CDS. The columns and their units are shown in Table \ref{tab:maindata}.
A continuously updated version of the catalog, as well as the scripts used for CASA processing, are available at \url{https://github.com/alvakinman/OSO_flux_monitoring}.

\begin{acknowledgements}
      This research was supported by the Swedish Research Council \emph{VR}, project
      number 2023-05185 (RISOTTO).
       This research has made use of the SIMBAD database, operated at CDS, Strasbourg, France.
      This research also made use of Astropy\footnote{\url{https://www.astropy.org}}:  a community-developed core Python package and an ecosystem of tools and resources for astronomy \citep{astropy2013, astropy2018, astropy2022}.
\end{acknowledgements}

\bibliography{references}
\bibliographystyle{aa}

\begin{appendix}
\nolinenumbers

\FloatBarrier
\twocolumn

\onecolumn

\newpage
\section{Light curves}
\label{app:sourceplots}
\begin{figure}[b!]
    \centering
    \includegraphics[width=0.95\linewidth,trim={0 3cm 0 3cm},clip]{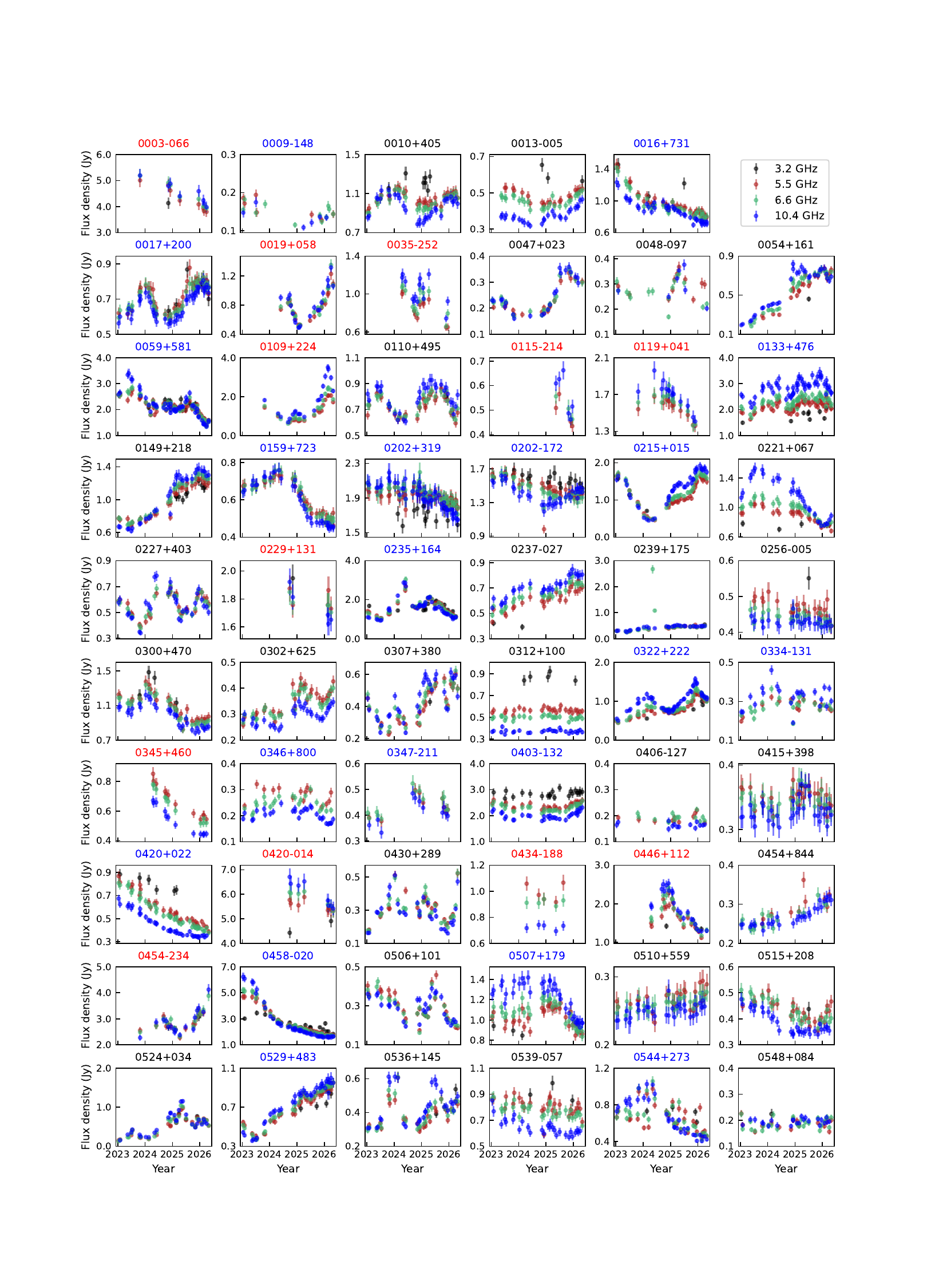}
    \caption{Flux density time series.}
    \label{fig:all_lightcurves1}
\end{figure}

\begin{figure}[]
    \centering
    \includegraphics[width=\linewidth,trim={0 3cm 0 3cm},clip]{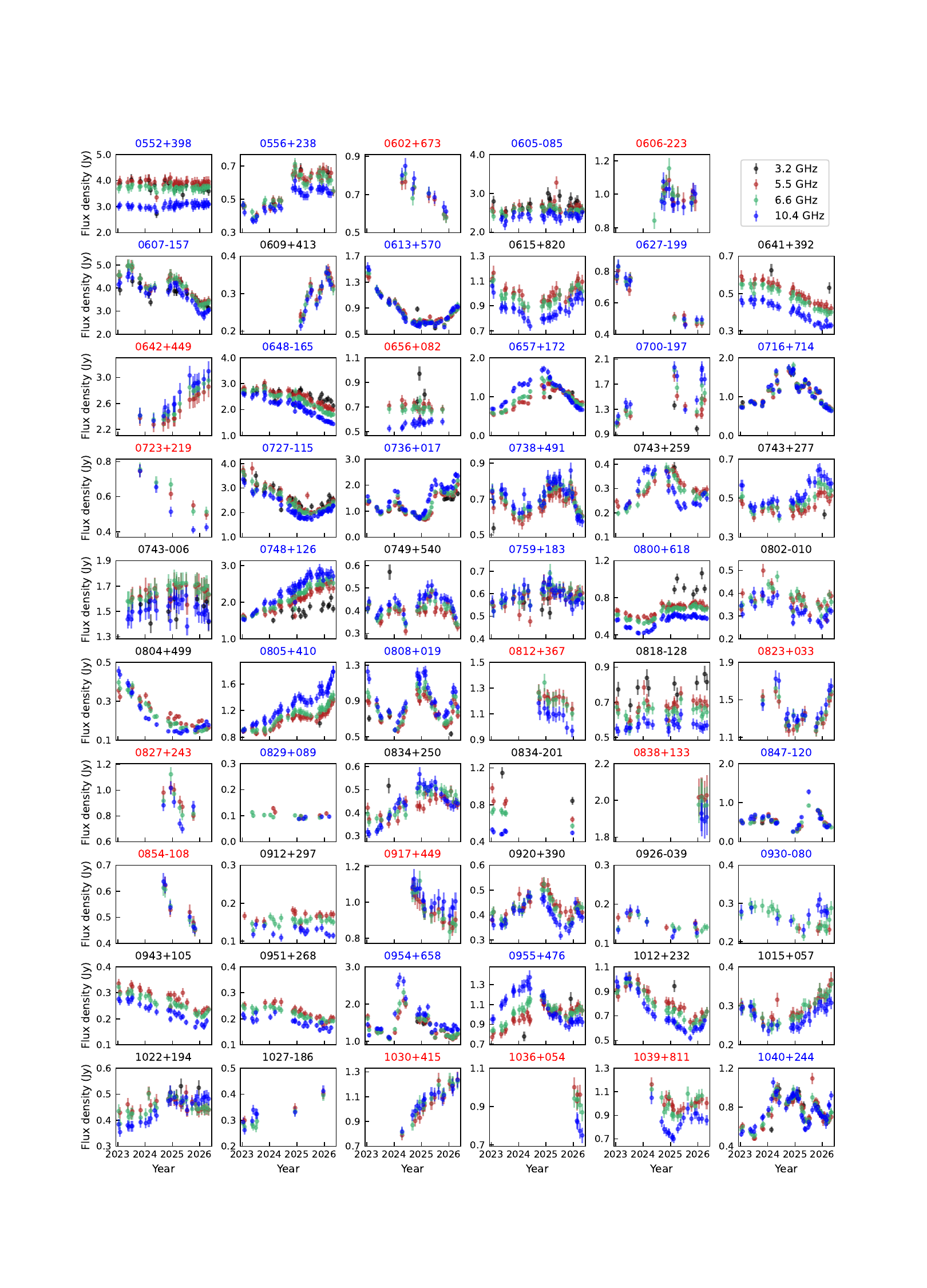}
    \caption{Flux density time series (continued).}
    \label{fig:all_ligthcurves2}
\end{figure}

\begin{figure}[]
    \centering
    \includegraphics[width=\linewidth,trim={0 3cm 0 3cm},clip]{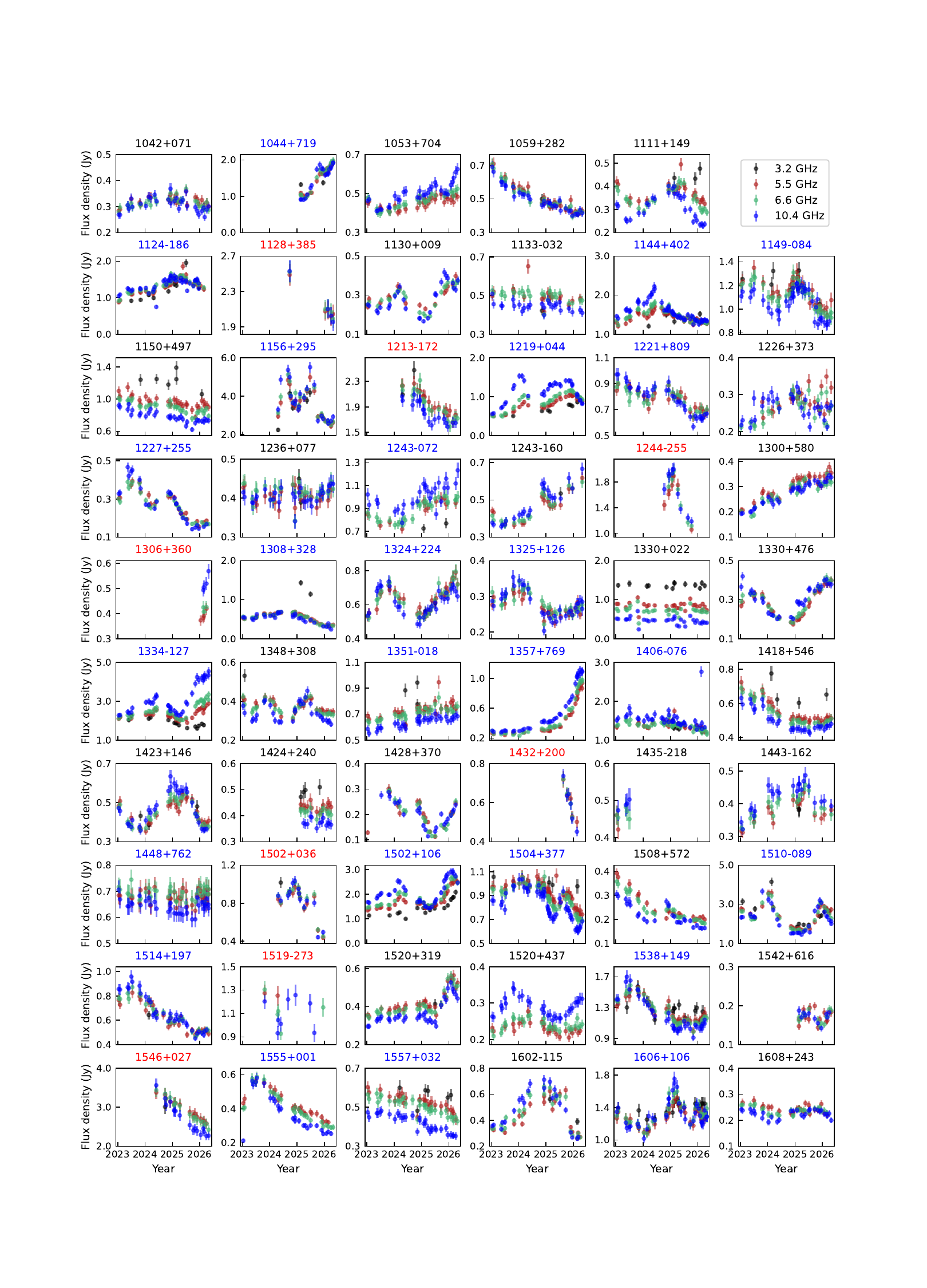}
    \caption{Flux density time series (continued).}
    \label{fig:all_lightcurves3}
\end{figure}

\begin{figure}[]
    \centering
    \includegraphics[width=\linewidth,trim={0 3cm 0 3cm},clip]{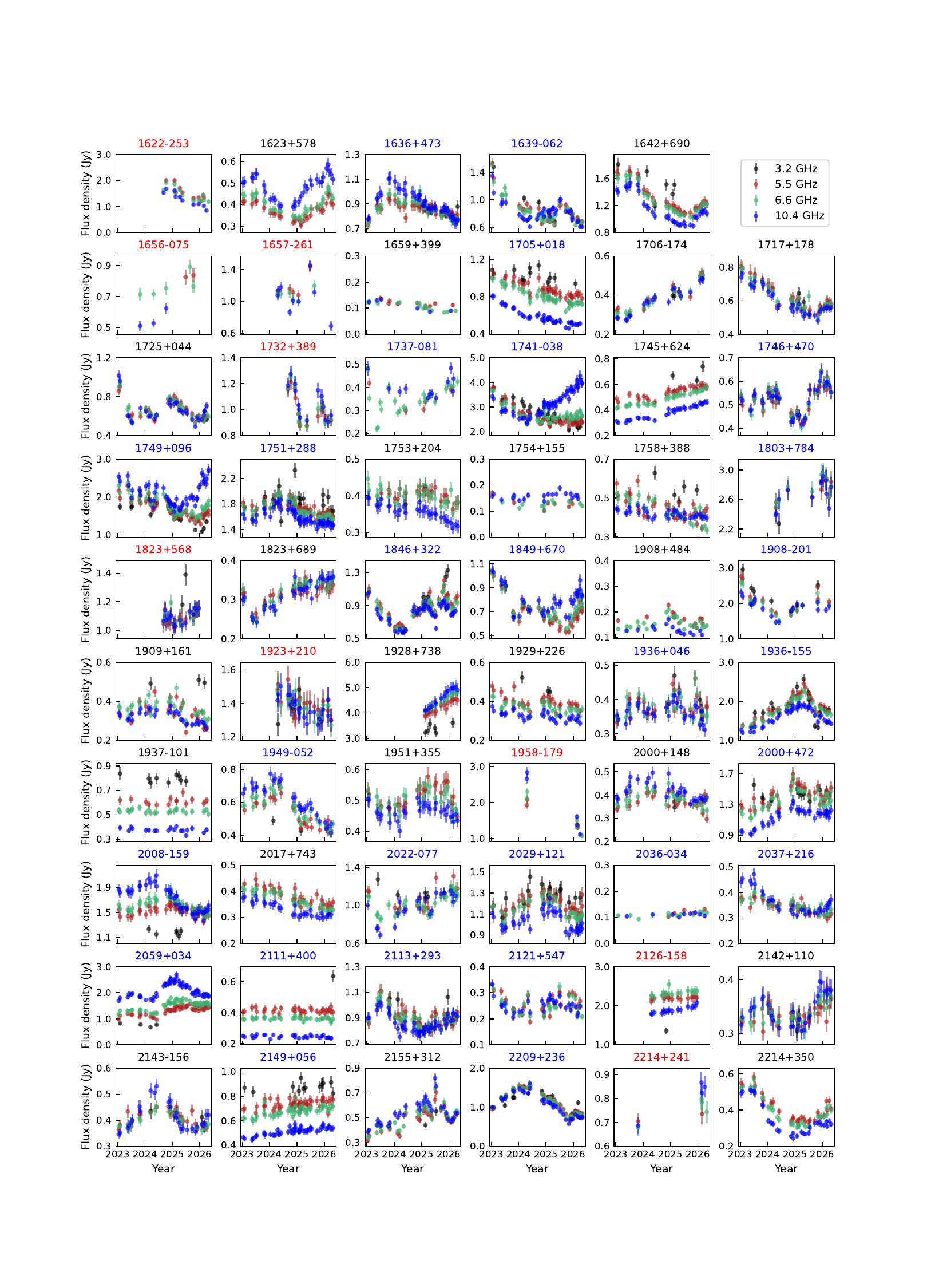}
    \caption{Flux density time series (continued).}
    \label{fig:all_lightcurves4}
\end{figure}

\begin{figure}[]
    \centering
    \includegraphics[width=1\linewidth]{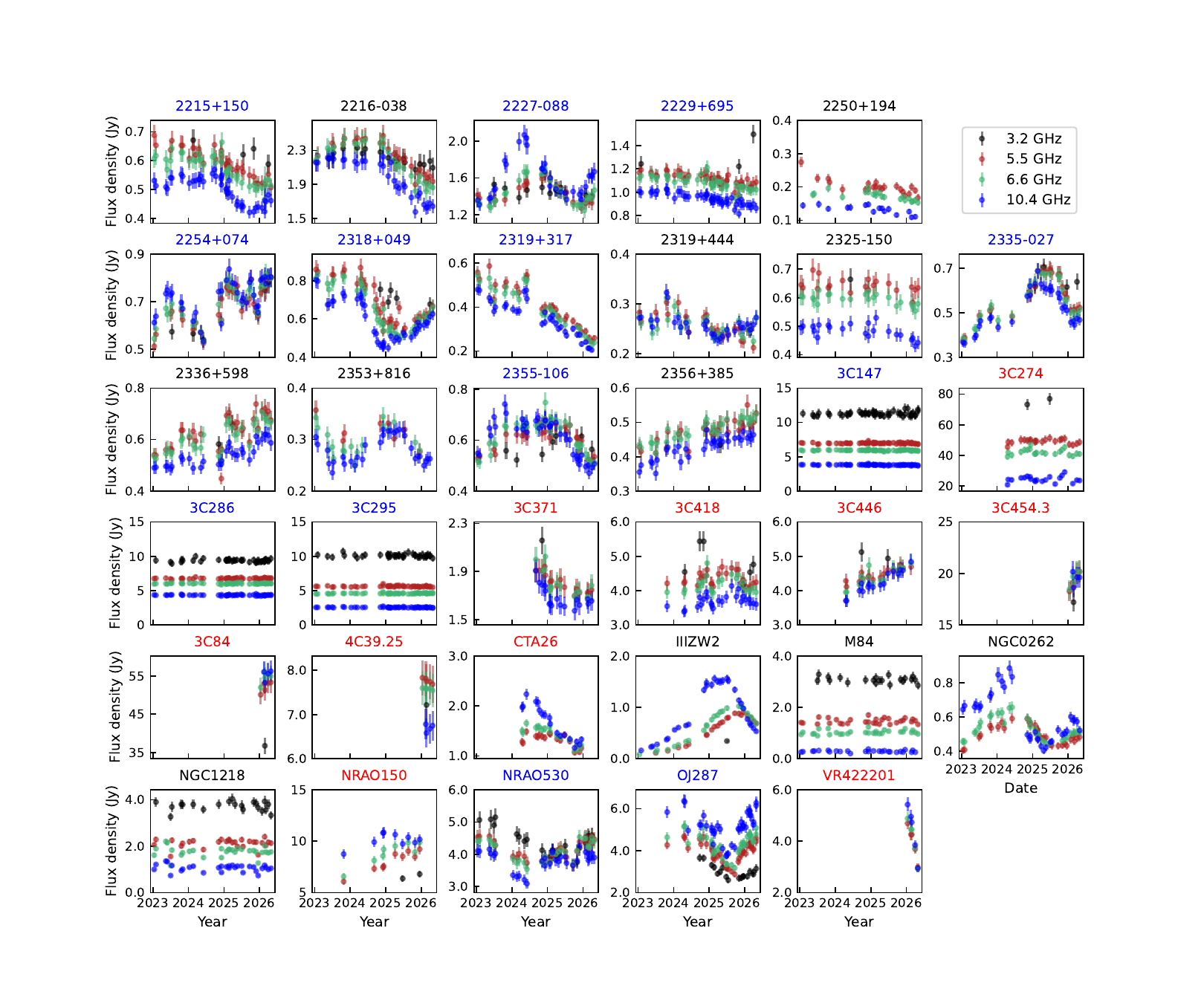}
    \caption{Flux density time series (continued).}
    \label{fig:all_lightcurves5}
\end{figure}

Fig.~\ref{fig:all_lightcurves1}-\ref{fig:all_lightcurves5} depict light curves of all sources with data from at least five experiments. 
The colors of the source names indicate the type of experiments that the sources have been observed in.
Black, red, and blue names represent only-FM, only-VO, and FM+VO, respectively. Sources are presented with names used by the IVS. For the majority of the sources, this corresponds to their B1950 coordinates and the designation used by the International Earth Rotation and Reference Systems Service (IERS). A translation table is provided in Tab.~\ref{tab:name-translation} for the sources where the IVS and IERS names are different. 

\begin{table} 
    \centering
    \caption{Translation table between IERS (B1950) source names and IVS names for the sources where they are different. The full table (including sources not shown in Fig.~\ref{fig:all_lightcurves1}--\ref{fig:all_lightcurves5}) is available at the CDS.}
    \small
    \begin{tabular}{ll|ll|ll|ll}
    \hline\hline
        IERS name & IVS name &  IERS name & IVS name &  IERS name & IVS name &  IERS name & IVS name\\
        \hline

0007+106 & IIIZW2 & 0538+498 & 3C147 & 1409+524 & 3C295 & 2223-052 & 3C446 \\ 
0046+316 & NGC0262 & 0851+202 & OJ287 & 1730-130 & NRAO530 & 2250+190 & 2250+194 \\ 
0305+039 & NGC1218 & 0923+392 & 4C39.25 & 1807+698 & 3C371 & 2251+158 & 3C454.3 \\ 
0316+413 & 3C84 & 1222+131 & M84 & 2017+745 & 2017+743 & & \\ 
0336-019 & CTA26 & 1228+126 & 3C274 & 2037+511 & 3C418 & & \\ 
0355+508 & NRAO150 & 1328+307 & 3C286 & 2200+420 & VR422201 & & \\
\hline

    \end{tabular}
    \label{tab:name-translation}
\end{table}

\FloatBarrier

In Tab.~\ref{tab:maindata}, the format of the catalog data file is exemplified. The full catalog is available at the CDS and on GitHub\footnote{\url{https://github.com/alvakinman/OSO_flux_monitoring}}. The numbers 1--4 correspond to VGOS bands A--D.

\begin{table*}[h!]
  \centering
        \caption{Sample data from flux density table. The full table is available at the CDS.}
        \label{tab:maindata}
        \small
\begin{tabular}{llllllllllll}

\hline\hline
source & time & date       & exp    & flux1 & flux2 & flux3  & flux4  & err1  & err2  & err3  & err4   \\
& MJD & YYYY-MM-DD      &    & Jy & Jy &  Jy &  Jy & Jy & Jy &  Jy & Jy \\
    \hline
2059+034 & 59958.5   & 2023-01-14 & fm3013 &            & 0.981      & 1.169      & 1.690       &           & 0.033     & 0.039     & 0.052     \\
2059+034 & 59976.4   & 2023-02-01 & fm3031 & 0.824      & 1.128      & 1.312      & 1.798      & 0.029     & 0.036     & 0.045     & 0.054     \\
2059+034 & 60079.1   & 2023-05-15 & fm3134 &            & 1.192      & 1.317      & 1.823      &           & 0.046     & 0.040      & 0.055     \\
2059+034 & 60093.1   & 2023-05-29 & fm3148 &            &            & 1.288      & 1.950       &           &           & 0.043     & 0.060      \\
2059+034 & 60127.0   & 2023-07-01 & fm3182 &            & 1.160       & 1.325      & 1.927      &           & 0.039     & 0.043     & 0.059     \\
2059+034 & 60140.0   & 2023-07-14 & fm3195 &            & 1.164      & 1.342      & 1.973      &           & 0.039     & 0.043     & 0.060      \\
2059+034 & 60237.7   & 2023-10-20 & fm3293 &            & 1.069      & 1.270       & 1.840       &           & 0.035     & 0.040      & 0.056     \\
2059+034 & 60244.2   & 2023-10-27 & vo3299 &            & 1.186      & 1.366      & 1.897      &           & 0.040      & 0.044     & 0.057     \\
2059+034 & 60250.8   & 2023-11-02 & fm3306 & 0.780       & 1.164      & 1.331      & 1.834      & 0.029     & 0.046     & 0.042     & 0.056     \\
2059+034 & 60321.0    & 2024-01-12 & fm4011 &            & 1.162      & 1.297      & 1.717      &           & 0.037     & 0.044     & 0.053     \\
 \hline
\end{tabular}
\end{table*}
\FloatBarrier

\section{List of experiments}
Tab.~\ref{app:exptable} contains the start and end time of all experiments used to produce the flux density catalog.

\small
\begin{longtable}{lll|lll}
\caption{List of analyzed experiments.}\\
\label{app:exptable}\\
\hline\hline
Experiment & Start time       & End time   & 
Experiment & Start time       & End time \\
\hline
\endfirsthead
\caption{continued.}\\
\hline
Experiment & Start time       & End time    & 
Experiment & Start time       & End time      \\
\hline
\endhead
\hline
\endfoot

fm3013     & 2023-01-13 21:00 & 2023-01-14 20:56 & fm5052     & 2025-02-21 14:00 & 2025-02-22 15:56 \\
fm3031     & 2023-01-31 23:00 & 2023-02-01 22:39 & fm5053     & 2025-02-22 17:00 & 2025-02-23 18:50 \\
fm3134     & 2023-05-14 10:00 & 2023-05-15 05:56 & fm5087     & 2025-03-28 18:00 & 2025-03-29 19:49 \\
fm3148     & 2023-05-28 10:00 & 2023-05-29 05:53 & vo5099     & 2025-04-09 00:00 & 2025-04-09 23:58 \\
fm3182     & 2023-07-01 10:37 & 2023-07-02 05:59 & fm5114     & 2025-04-24 09:00 & 2025-04-25 10:38 \\
fm3195     & 2023-07-14 17:30 & 2023-07-15 13:25 & vo5134     & 2025-05-14 00:00 & 2025-05-15 00:00 \\
fm3293     & 2023-10-20 12:00 & 2023-10-21 07:38 & fm5139     & 2025-05-19 09:00 & 2025-05-20 10:52 \\
vo3299     & 2023-10-26 18:00 & 2023-10-27 18:00 & vo5176     & 2025-06-25 00:00 & 2025-06-26 00:00 \\
fm3306     & 2023-11-02 13:30 & 2023-11-03 09:16 & fm5183     & 2025-07-02 10:00 & 2025-07-03 11:53 \\
fm4011     & 2024-01-11 15:30 & 2024-01-12 11:30 & fm5197     & 2025-07-16 10:00 & 2025-07-17 11:59 \\
fm4050     & 2024-02-19 13:30 & 2024-02-20 09:29 & vo5232     & 2025-08-20 00:00 & 2025-08-21 00:00 \\
fm4074     & 2024-03-14 15:00 & 2024-03-15 10:54 & fm5264     & 2025-09-22 08:30 & 2025-09-22 11:30 \\
vo4108     & 2024-04-17 12:02 & 2024-04-18 11:57 & vo5281     & 2025-10-08 00:00 & 2025-10-09 00:00 \\
vo4115     & 2024-04-24 12:00 & 2024-04-25 11:58 & vo5302     & 2025-10-29 00:00 & 2025-10-29 23:59 \\
fm4128     & 2024-05-07 08:00 & 2024-05-08 05:37 & fm5306     & 2025-11-02 14:00 & 2025-11-03 15:55 \\
vo4150     & 2024-05-29 12:00 & 2024-05-30 11:57 & vo5330     & 2025-11-26 07:01 & 2025-11-27 00:00 \\
fm4155     & 2024-06-03 16:00 & 2024-06-04 12:00 & fm5331     & 2025-11-27 07:30 & 2025-11-28 09:19 \\
vo4248     & 2024-09-04 12:00 & 2024-09-05 12:00 & vo5351     & 2025-12-17 00:00 & 2025-12-18 00:00 \\
vo4269     & 2024-09-25 12:00 & 2024-09-26 12:00 & fm5352     & 2025-12-18 06:00 & 2025-12-19 08:00 \\
vo4283     & 2024-10-09 12:00 & 2024-10-10 11:57 & vo6014     & 2026-01-14 00:00 & 2026-01-15 00:00 \\
vo4311     & 2024-11-06 12:00 & 2024-11-07 12:00 & fm6029     & 2026-01-29 04:00 & 2026-01-30 05:50 \\
fm4312     & 2024-11-07 15:00 & 2024-11-08 16:43 & vo6049     & 2026-02-18 00:00 & 2026-02-19 00:00 \\
vo4339     & 2024-12-04 12:05 & 2024-12-05 12:00 & fm6054     & 2026-02-23 18:00 & 2026-02-24 19:59 \\
fm4340     & 2024-12-05 13:00 & 2024-12-06 14:53 & vo6056     & 2026-02-25 00:00 & 2026-02-25 23:59 \\
fm4341     & 2024-12-06 17:00 & 2024-12-07 18:52 & fm6066     & 2026-03-07 02:00 & 2026-03-08 03:59 \\
vo4346     & 2024-12-11 12:00 & 2024-12-12 11:59 & vo6091     & 2026-04-01 00:00 & 2026-04-02 00:00 \\
vo5022     & 2025-01-22 00:00 & 2025-01-22 23:59 & fm6092     & 2026-04-02 02:00 & 2026-04-03 03:59 \\
fm5027     & 2025-01-27 10:00 & 2025-01-28 11:24 & vo6119     & 2026-04-29 00:00 & 2026-04-30 00:00 \\
vo5029     & 2025-01-29 00:03 & 2025-01-30 00:00 & fm6120     & 2026-04-30 02:00 & 2026-05-01 03:59 \\
vo5043     & 2025-02-12 00:00 & 2025-02-13 00:00 & & &

\end{longtable}
\normalsize

\FloatBarrier

\newpage

\section{Comparison of outlier filtering methods}
\label{app:outlier}
In this section we compare the number of data points that are flagged during outlier filtration, depending on which method is used. We compare the cross-polarization criterion from \cite{varenius2022} and \cite{kinman2025} to the deviation-from-median criterion used in this work (see Section \ref{met:outliers}). 

\begin{figure*}[h!]
\begin{subfigure}{0.49\linewidth}
    \centering
    \includegraphics[width=\linewidth]{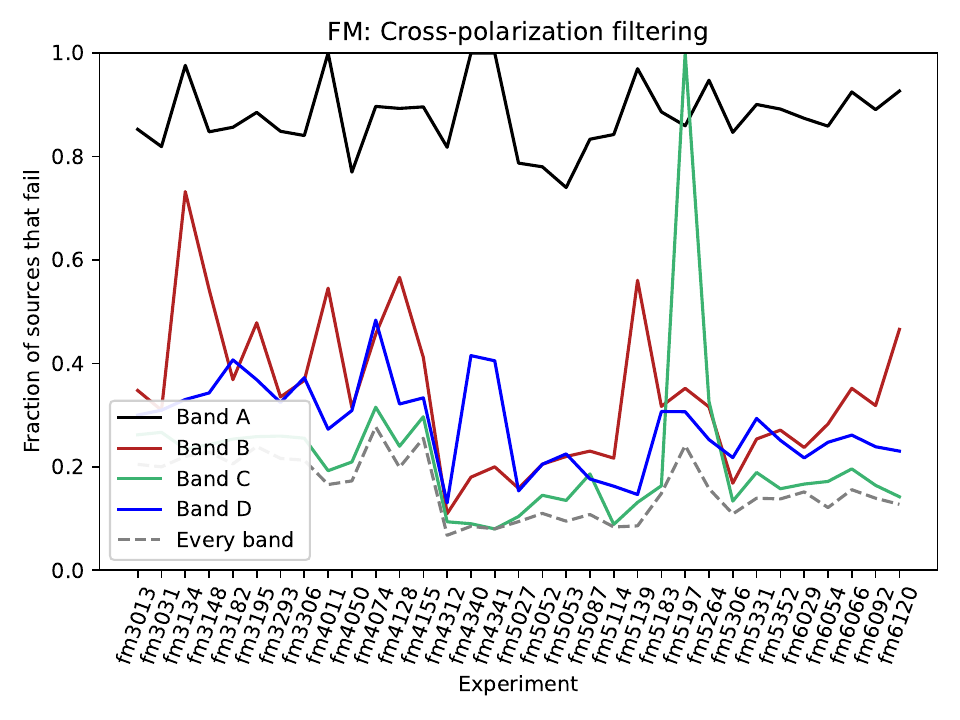}
    \caption{cross-polarization criterion}
    \label{fig:enter-label}
\end{subfigure}
\begin{subfigure}{0.49\linewidth}
    \centering
    \includegraphics[width=\linewidth]{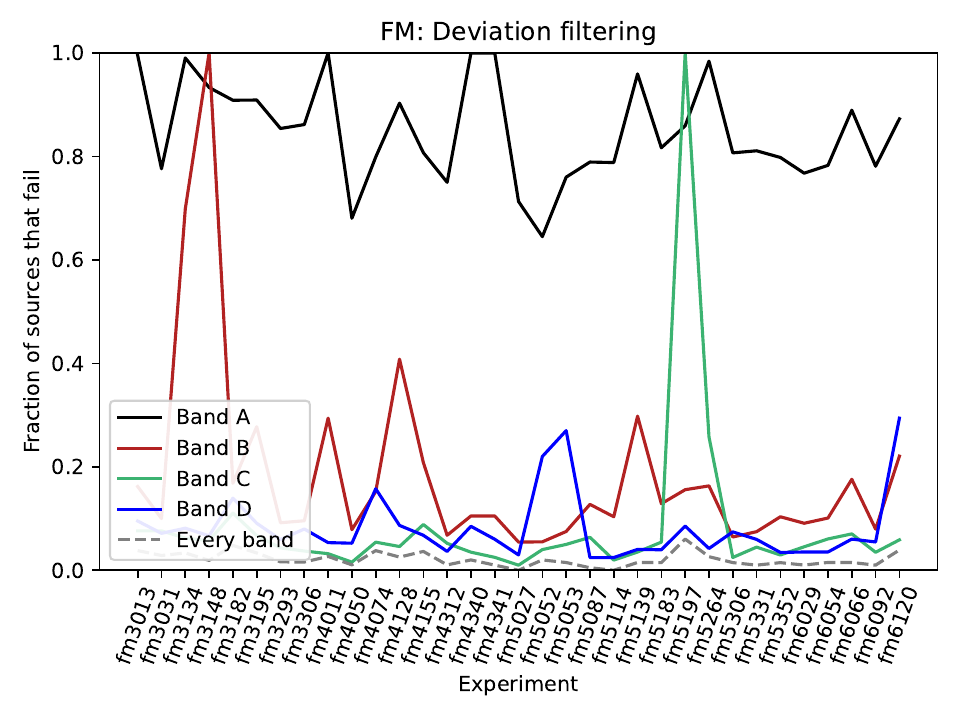}
    \caption{deviation criterion}
    \label{fig:enter-label}
\end{subfigure}
    \caption{Fraction of failed sources in each FM experiment, with the old cross-polarization criterion (a) and the new deviation criterion (b). Note: band C was observed with an incorrect frequency setup in fm5197, so all band C data are removed. Band A was not observed in fm4340 and fm4341.}
    \label{fig:failrate-fm}
\end{figure*}

\begin{figure*}[h!]
\begin{subfigure}{0.49\linewidth}
    \centering
    \includegraphics[width=\linewidth]{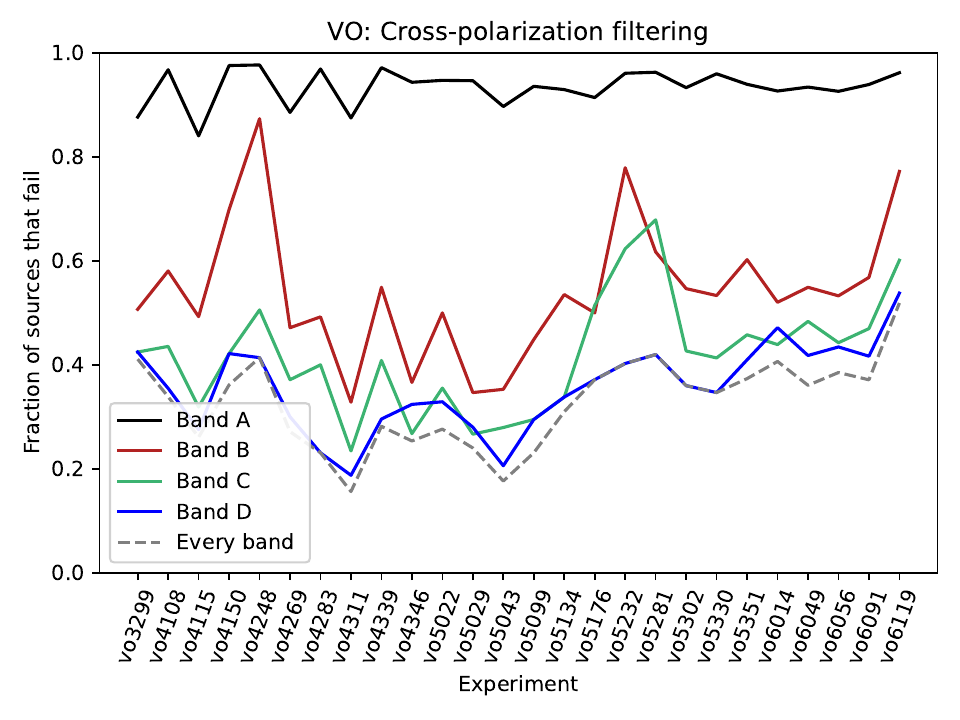}
    \caption{cross-polarization criterion}
    \label{fig:enter-label}
\end{subfigure}
\begin{subfigure}{0.49\linewidth}
    \centering
    \includegraphics[width=\linewidth]{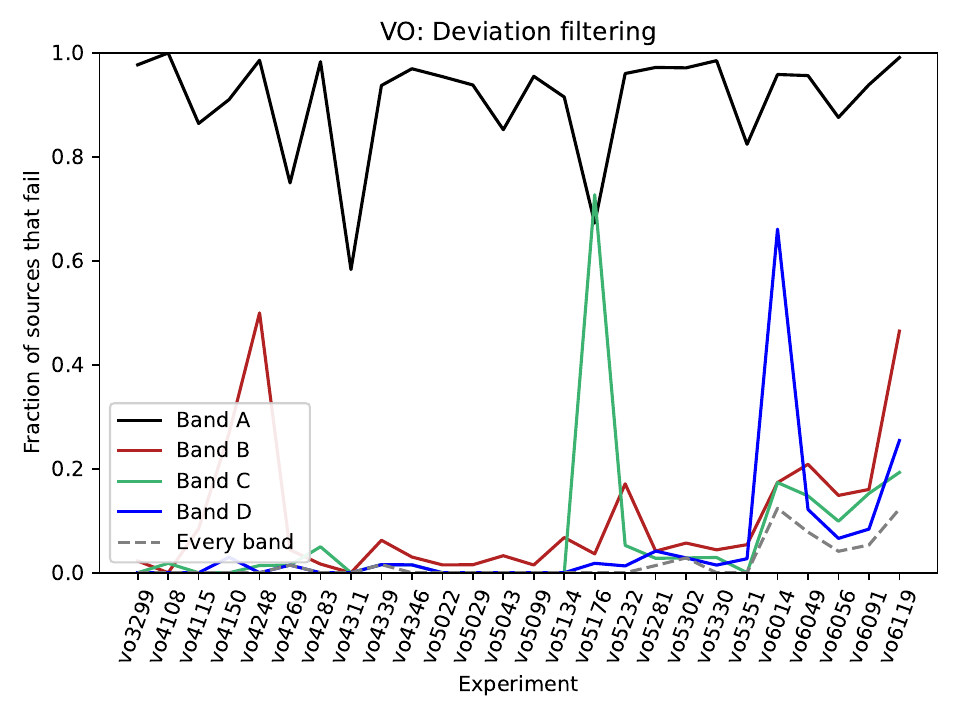}
    \caption{deviation criterion}
    \label{fig:enter-label}
\end{subfigure}
    \caption{Fraction of failed sources in each VO experiment, with the old cross-polarization criterion (a) and the new deviation criterion (b).}
    \label{fig:failrate-vo}
\end{figure*}

Fig.~\ref{fig:failrate-fm}-\ref{fig:failrate-vo} show the fraction of data points that were flagged in each band. This includes only the sources where no final flux density value was obtained for the band due to the majority of the sub-bands being flagged. For both FM and VO experiments, we find that the deviation criterion keeps more data.

\FloatBarrier

\newpage

\section{Correlation between flux density calibrator sources}
\label{app:correlation}
\begin{figure*}[h!]
\begin{subfigure}{0.49\linewidth}
    \centering
    \includegraphics[width=\linewidth]{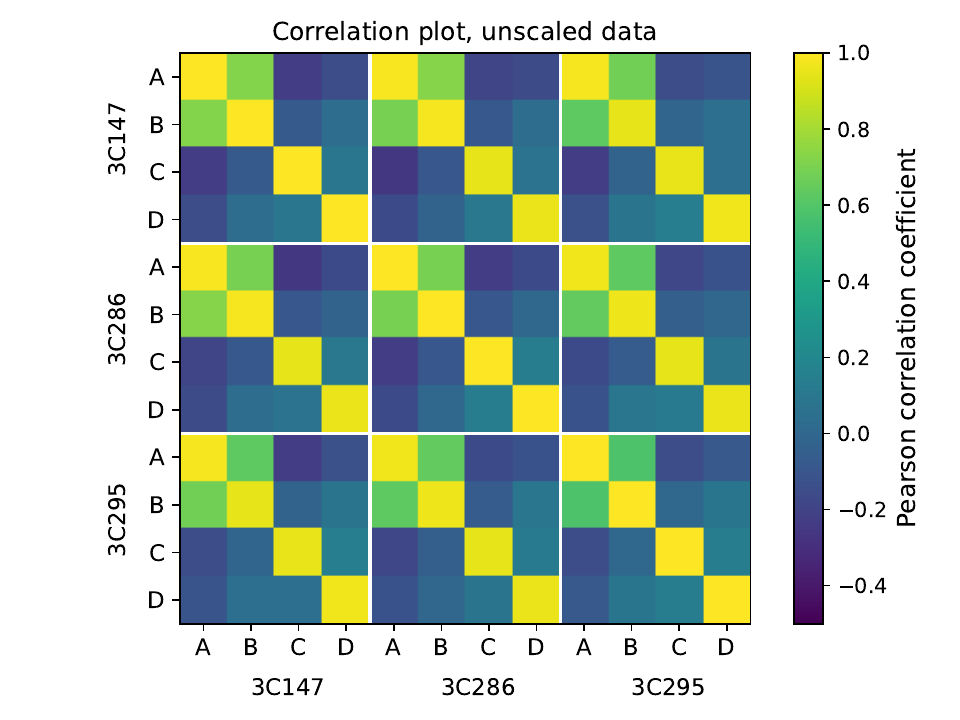}
    \caption{before applying scale factors}
    \label{fig:correlation_bigplot1}
\end{subfigure}
\begin{subfigure}{0.49\linewidth}
    \centering
    \includegraphics[width=\linewidth]{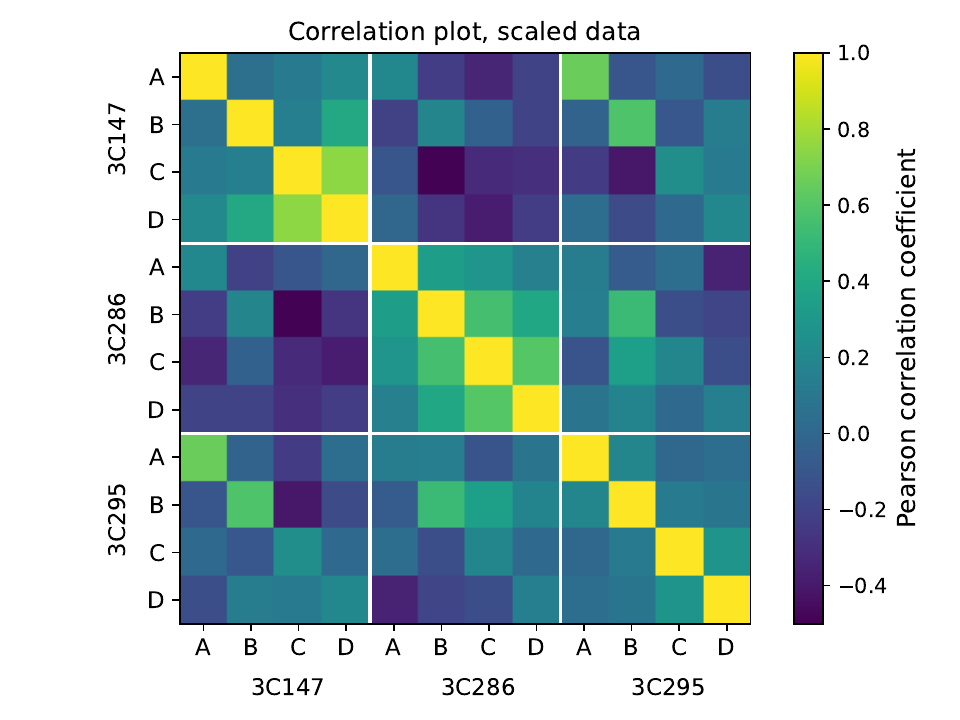}
    \caption{after applying scale factors}
     \label{fig:correlation_bigplot2}
\end{subfigure}
    \caption{Pearson correlation coefficients for all calibrator sources, before (a) and after (b) scaling. For example, the value in the intersection of "3C147 A" and "3C286 D" shows the correlation between the flux density of 3C147 in band A and the flux density of 3C286 in band D.  Before scaling, each source is strongly correlated with the other sources in the same band. There is also a clear correlation between band A and B for the same source. After scaling, these systematic effects are reduced.}
    \label{fig:correlation_bigplot}
\end{figure*}

Fig.~\ref{fig:correlation_bigplot} shows the Pearson correlation coefficient between all calibrator sources and bands, before and after scaling. Before scaling, the correlation between different sources is above 0.9. There is also a high correlation between band A and B of the same source. After scaling, these effects are reduced.

\section{Bandpass S/N factor}
\label{app:bandpass}
\begin{figure}[b!]
    \centering
    \includegraphics[width=0.55\linewidth]{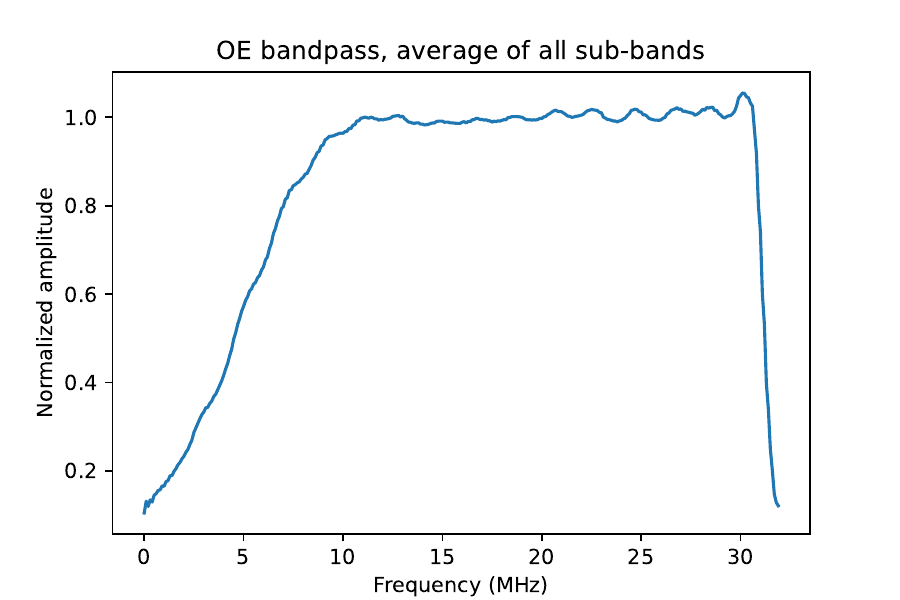}
    \caption{Bandpass shape averaged over all sub-bands, obtained from the DBBC3 of ONSA13NE during experiment vo5022.}
    \label{fig:bandpass}
\end{figure}

The theoretical S/N is computed under the assumption of a 32~MHz sub-band bandwidth. However, the filters of the DBBC3, used in the OTT signal chain, do not have a perfect rectangular bandpass. As described in \cite{mccallum2022}, the taper of the bandpass response gives an effective bandwidth that is lower than 32~MHz. The effective bandwidth can be obtained by integrating the bandpass curve, shown for ONSA13NE in Fig.~\ref{fig:bandpass}. This results in a value of 26.8~MHz. Since the S/N equation contains the bandwidth under a square root, the resulting bandpass efficiency factor is  $\sqrt{26.8/32} \approx 0.92$.

\end{appendix}
\end{document}